\documentclass[a4paper,fleqn,usenatbib,useAMS]{mnras}

\usepackage{graphicx}	
\usepackage{amsmath}	
\usepackage{amssymb}	
\usepackage{multicol}        
\usepackage{bm}		
\usepackage{pdflscape}	
\usepackage{multirow}

\usepackage{color}
\newcommand{\Msun}{{\rm ~M}_{\odot}}

\newcommand{\Zsun}{{\rm ~Z}_{\odot}}

\usepackage[T1]{fontenc}
\usepackage{ae,aecompl}

\usepackage{txfonts}

\title[Metallicity of the star formation]{Metallicity of the star formation based on the observational
properties of star forming galaxies}

\title[Metallicity of stars throughout the cosmic history]{Metallicity of stars formed throughout the cosmic history based on the observational
properties of star forming galaxies}

\author[M. Chruslinska et al.]
{Martyna Chruslinska,$^{1}$\thanks{E-mail: m.chruslinska@astro.ru.nl}
Gijs Nelemans$^{1,2}$
\\
$^{1}$Department of Astrophysics/IMAPP, Radboud University, P O Box 9010, NL-6500 GL Nijmegen, The Netherlands\\
$^{2}$Institute for Astronomy, KU Leuven, Celestijnenlaan 200D, 3001 Leuven, Belgium
}
\date{Last updated 2019 May 22; in original form 2019 May 1}

\pubyear{2019}
\begin{document}
\label{firstpage}
\pagerange{\pageref{firstpage}--\pageref{lastpage}}
\maketitle

\begin{abstract}
Metallicity is one of the crucial factors that determine stellar evolution.
To characterize the properties of stellar populations
one needs to know the fraction of stars forming at different metallicities.
Knowing how this fraction evolves over time is necessary e.g.
to estimate the rates of occurrence of any stellar evolution related phenomena 
(e.g. double compact object mergers, gamma ray bursts).
Such theoretical estimates can be confronted with observational limits to validate 
the assumptions about the evolution of the progenitor system leading to a certain transient.
However, to perform the comparison correctly one needs to know the uncertainties related to the assumed
star formation history and chemical evolution of the Universe. 
\\
We combine the empirical scaling relations and other observational properties
of the star forming galaxies to construct the distribution of 
the cosmic star formation rate density at different metallicities and redshifts.
We address the question of uncertainty of this distribution due to currently unresolved questions, 
such as the absolute metallicity scale, the flattening in the star formation$-$mass relation or
the low mass end of the galaxy mass function.
We find that the fraction of stellar mass formed at metallicities <10\% solar (>solar) since z=3 varies
by $\sim$18\% ($\sim$26\%) between the extreme cases considered in our study.
This uncertainty stems primarily from the differences in the mass metallicity relations
obtained with different methods. 
We confront our results with the local core-collapse supernovae observations.
Our model is publicly available.
\end{abstract}




\section{Introduction}
The distribution of the star formation over metallicities varies 
throughout the history of the Universe.
It is an important ingredient to estimate
the rate of occurrence of any stellar or binary evolution related phenomena,
such as different types of supernovae, double compact object mergers or gamma ray bursts
\citep[e.g.][]{LangerNorman06,Dominik13,Belczynski16N,Mapelli17,Chruslinska19,Eldridge19}.
Such theoretical estimates can be, for instance, confronted with observational determinations of rates
to validate the assumptions about the evolution of the progenitor star or system
leading to a certain type of transient. 
However, to perform the comparison correctly, one needs to know the uncertainties related to the
assumed star formation history and chemical evolution of the Universe. 
This is especially important in the case of transients whose
formation scenarios are particularly sensitive to metallicity, e.g. long gamma ray bursts (long GRB)
and stellar double black hole mergers, 
both forming much more efficiently at low metallicities 
($\lesssim$0.1 solar metallicity; see e.g. \citet{LangerNorman06}, \citet{Stanek06}, 
\citet{WoosleyHeger06}, \citet{Palmerio19}
for long GRB and e.g. \citet{Belczynski10}, \citet{EldridgeStanway16}, \citet{Stevenson17},
\citet{Giacobbo18}, \citet{Klencki18} 
for double black hole mergers).
\\
Recently, \citet{Chruslinska19} 
demonstrated that the merger rate density
of double black holes estimated for the same description of the evolution of their progenitor systems
can be significantly different (even by a factor of $\sim$10) depending on the assumed
distribution of the cosmic star formation
rate density at different metallicities and time (probed by redshift; $SFRD(Z,z)$). 
This underlines the need for a better constrained picture
of the star formation history and chemical evolution of the Universe and of
understanding the associated uncertainties.
\\
So far different groups have taken different approaches to determine the $SFRD(Z,z)$,
often combining observations, theoretical inferences and/or cosmological simulations
 \citep[e.g.][]{LangerNorman06,Niino11,Dominik13,Belczynski16N,Lamberts16,Mapelli17,Schneider17}.
\\
In this paper we combine the empirical scaling relations from various observational studies 
describing the properties of star forming galaxies (star formation rate, mass, metallicity)
to construct the distribution of the cosmic star formation rate density at different
metallicities and time/redshift. 
Our method is outlined in Sec. \ref{sec:method}.
We also address the question of the uncertainty of this distribution, given
the currently unresolved problems (discussed in detail in Sec. \ref{sec: section2}), 
such as the absolute metallicity scale \citep[e.g.][]{KewleyEllison08} or the flattening (or lack of it)
at the high mass part of the star formation mass relation \citep[e.g.][]{Speagle14,Lee15}.
Systematic evaluation of these uncertainties in the context of $SFRD(Z,z)$ is still lacking in the literature.
\\
Our results are summarized in Sec. \ref{sec: results}.
In Sec. \ref{sec: CCSN rates} we apply them to calculate the volumetric core-collapse supernovae
rate and their local rate as a function of metallicity and contrast those quantities with observations.
We discuss the reliability of our results at high redshifts 
and present a brief comparison with simulations and earlier studies in Sec. \ref{sec: discussion}.
\\
The results of our calculations are publicly available at \url{https://ftp.science.ru.nl/astro/mchruslinska/}.
They can be applied to calculate the cosmological rates of various stellar evolution related
events and to asses their uncertainty due to uncertainties in observationally inferred $SFRD(Z,z)$.
They can also be contrasted with the results from cosmological simulations.
\\
Where appropriate we adopt a standard flat cosmology with $\Omega_{M}$=0.3, $\Omega_{\Lambda}$=0.7
and H$_{0} =70 \rm \ km \ s^{-1} \ Mpc^{-1}$ and assume a \citet{Kroupa01} initial mass function (IMF).

\section{Properties of star forming galaxies}\label{sec: section2}
While the chemical and star formation histories of individual galaxies are
heavily dependent on their environment and merger history, the picture
emerging from large galaxy surveys suggests that when a sufficiently large volume is considered,
the average properties of star forming galaxies follow relatively tight and simple, power-law like relations.
At a certain redshift both the star formation rate and metallicity correlate with the stellar masses
of galaxies \citep[e.g.][]{Brinchmann04,Tremonti04}.
Those correlations span many orders of magnitude in mass and are present in the entire
redshift range where observations are available.
\\
Below we provide an overview of the current observational results, 
underlining the open questions and uncertainties concerning the determination of various properties of star forming galaxies.
Based on the information presented in this section we decide on the relations and parameters used in our model.
Our choices are summarized below the relevant paragraphs.

\subsection{ Galaxy stellar mass function of star forming galaxies }\label{sec: intro:GSMF}

The mass distribution of star forming galaxies can be inferred from the galaxy stellar mass function 
(GSMF, the number density of galaxies per logarithmic mass bin). 
Observationally such estimates have been provided by many groups
focusing on different redshift ranges:
e.g \citet{Baldry04}, \citet{Baldry12}, \citet{Weigel16} at z$\sim$0 ,
\citet{Fontana06}, \citet{Ilbert13}, \citet{Moustakas13},\citet{Muzzin13},\citet{Tomczak14},\citet{Davidzon17}
at intermediate and high redshifts.  
There are several studies aiming to determine 
the GSMF at redshifts as high z$\gtrsim$7 \citep[e.g.][]{Duncan14,Grazian15,Stefanon15}
or even z$\gtrsim$8 \citep{Song16,Bhatawdekar18}.
\\
GSMF estimates for the same redshift bin can vary
substantially between different studies \citep[e.g. see figure 1 in][
comparing best fit GSMF from different studies]{Conselice16}, 
especially at the high and low mass end. 
Furthermore, the transition between redshift bins connecting results from different surveys
in general is not smooth, producing artificial jumps in the projected evolution of the GSMF over the cosmic history.
There is a number of factors that may be responsible for those differences.
\\
For instance, stellar mass estimates usually rely on stellar population synthesis model fits to the
measured SEDs of galaxies. The result of SED-fitting depends on 
the parameters of the model (stellar population models, metallicity, dust law),
star formation histories and IMF used.
Different assumptions lead to systematic offsets in the mass estimates
\citep[e.g.][]{Marchesini09}. 
Furthermore, GSMF estimates are based on different surveys, focusing on different redshift ranges (and wavelengths),
varying in depth and width of the covered sky area, completeness limits and introducing different biases.
On top of that, in case of GSMF for the star forming/quiescent subsample of galaxies
the final result is affected by the criteria used to select the interesting
population \citep[e.g.][]{Baldry12}. This choice is usually based on the colour $-$ colour diagram, 
with various color indices and selection criteria used.
\\
The observed (either for the active or total sample) galaxy stellar mass function generally declines with mass, 
shows a sharp cut-off at high masses (around $M_{\ast}\sim10^{10.7} \Msun$ at z$\sim$0) 
and a power-law tail at low masses,
a relation well described and commonly fitted with a \citet{schechter76}(or sometimes double Schechter) 
function:
\begin{equation}
 n_{gal}=\Phi(M_{*}) \ \rm d \ M_{*} = \Phi_{*} e^{-M/M_{*}} 
	\left( \frac{M}{M_{cut \ off}} \right)^{\alpha_{GSMF}} dM_{*}
\end{equation}
where $n_{gal}$ is the number density of galaxies in a mass bin $d M_{*}$,
$M_{\rm cut\ off}$ is the stellar mass at which the Schechter function bends, 
departing from a single power law with 
slope $\alpha_{\rm GSMF}$ at low masses to an exponential cut off at high masses. 
$\Phi_{*}$ provides the normalization (number density at $M_{\rm cut\ off}$).
The slope of the low mass end of GSMF is particularly weakly constrained. 
Even though the low mass galaxies are expected to be the most abundant in the Universe, 
they are also faint and difficult to observe especially at higher $z$. 
Also the mass completeness limit of any deep survey increases with redshift
(e.g. the sample from ZFOURGE survey used by \citet{Tomczak14} at z$\sim$0.3 is 
complete down to $M_{\ast}\sim10^{8} \Msun$, 
while at z$\sim$3 the completeness limit moves to $M_{\ast}\gtrsim 10^{9} \Msun$).
This leaves the GSMF in the low mass dwarf galaxy regime unconstrained.\\
Taking the fitted low mass slopes from different studies (see Tab. \ref{tab: GSMF}) at face value,
one would conclude that there is an overall tendency for the slope to steepen with $z$ 
($\alpha_{\rm GSMF}$ becomes more negative).
However, keeping in mind the differences in methods and surveys used by different authors,
correlations between the parameters of the fits 
\citep[in particular between $\alpha_{\rm GSMF}$ and $M_{\rm cut\ off}$, e.g.][]{Grazian15,Song16,Weigel16}
and the fact that the source of discrepancies between the results from different studies for a single redshift
bin is not well understood, such a simple comparison may not reflect a true evolution of $\alpha_{\rm GSMF}$.
Indeed, while some authors find evidence for $\alpha_{\rm GSMF}$ getting more negative with increasing $z$
\citep[e.g.][]{Ilbert13,Song16,Bhatawdekar18},
such evolution is not always found within one study covering a range of redshifts
\citep[e.g.][]{Marchesini09,Duncan14,Tomczak14,Grazian15}. 
\\
\newline
\underline{Our choice:} Instead of following the results of one group
we opt to average the estimates provided by different authors, 
similarly to the approach taken by \citet{Henriques15} to constrain
their semi-analytic model.
\\
The weakly constrained low mass end slope of the GSMF is treated separately.
We allow for two variations: one in which this slope is fixed to $\alpha_{fix}$=-1.45 
as found by \citet{Baldry12} (this seems to be a good compromise at least between z=0
and z=2 -- during the bulk of the cosmic history, see Fig. \ref{fig: alpha_GSMF_vs_z}) and one in which it increases
with redshift, following the linear fit shown in Fig. \ref{fig: alpha_GSMF_vs_z}. 
See Sec. \ref{sec: method: GSMF} for the details.

\subsection{Metallicity} \label{sec: metallicity: intro}

In this paper we use the word `metallicity` in a general sense 
(as a measure of the abundance of elements heavier than helium)
and use the symbol $Z_{O/H}$ (or write explicitly $12 + \rm log \left(O/H \right)$)
when referring specifically to the oxygen abundance ratio 
and $Z$ when referring specifically to the mass fraction of heavy elements.
\\ \newline
The observations that we use provide estimates of metallicity in terms of oxygen abundance 
$Z_{O/H}$. 
However, the metal mass fraction is perhaps more useful for practical applications of our model
and hence we also present our results in terms of $Z$.
To convert $Z_{O/H}$ to $Z$, we assume a simple scaling of the metal abundances with
$Z_{O/H}$ that maintains the solar abundance ratios 
(i.e. $log\left(Z/Z_{\odot}\right) = Z_{O/H} - Z_{O/H \odot}$).
\\
However, there is little consensus in the literature regarding the value of
solar metallicity and solar composition.
Throughout the paper we assume solar abundances found by \citet{AndersGrevesse89} 
 ($Z_{O/H \odot}$=8.83 and $Z_{\odot}$=0.017), 
 since their results fall roughly in the middle of the range of the presently reported values
 (see appendix \ref{app: nomenclature}).
 All estimates of $Z$ shown within our results were calculated assuming these values.
 We stress that this conversion is not unique and for the reasons discussed
 below in Sec. \ref{sec: O vs Fe} should be taken with caution.

\subsubsection{Oxygen vs iron abundance}\label{sec: O vs Fe}
The strong metallicity dependence of the efficiency of the
formation of various transients originating from massive stars 
(e.g. long gamma ray bursts, double black hole mergers)
is primarily driven by the abundance of iron.
This is because stellar winds from massive O$-$type and Wolf Rayet stars are driven 
by the radiation pressure on metal lines, 
and Fe easily dominates the atmospheric opacity due to its complex atomic structure
\citep[e.g.][]{Pauldrach86,Vink01,VinkDeKoter05,Vink11}.\\
Low wind mass loss rates are for instance necessary for a star to maintain high angular momentum,
which is needed to produce a long gamma ray burst \citep[e.g.][]{WoosleyHeger06}.
At the same time, lower wind mass loss allows for the formation of a more massive BH progenitor,
which then may form a BH in a direct collapse 
\citep[with no mass loss and small/no natal kick
\footnote{Unless asymmetric neutrino emission during the collapse can produce substantial
BH natal kicks \citep[e.g.][]{FryerKusenko06}.}
; e.g.][]{FryerKalogera01,Fryer12}.
The binary containing such a BH likely remains bound after its formation
and may evolve towards a merging double black hole system \citep[e.g.][]{Klencki18}.
\\
Hence, it would be a more natural choice for our study to consider Fe instead of O abundances.
However, observational determination of the iron abundance is challenging and the number of
available results present in the literature much more limited than in the case of oxygen.
We thus rely on $Z_{O/H}$ measurements assuming that it provides a good representation
of the overall metallicity (in particular the iron abundance) in the star forming material.
\\
This is an important simplification to keep in mind, as the relative abundances of O and Fe in general do not
follow a simple proportionality, notably because 
the interstellar medium is enriched with different elements 
on different timescales \citep[e.g.][]{WheelerSnedenTruran89}. 
While oxygen is generously released by massive stars (and hence on short timescales
$\sim$10 Myr), a significant fraction of iron enrichment occurs via type Ia supernovae
(over much longer $\sim$Gyr timescales). 
As a result, young star forming systems often reveal an overabundance of oxygen relative to iron 
with respect to solar \citep[e.g.][]{ZhangZhao05,Izotov06}.
The different timescales at which O and Fe abundance evolve in the interstellar medium
are reflected in the [Fe/O] vs [Fe/H] (or [O/H]) 
relation that can be obtained for a given stellar system \citep[e.g.][]{Tolstoy09}. 
The relation provides a better way to translate the O to Fe abundance for a particular system,
but is not universal (as it depends e.g. on the star formation history and the IMF) 
and hence is not applicable to our study.
Top-heavy IMF is another factor that can lead to increased ratio of 
$\alpha-$elements to iron \citep[e.g.][]{Hashimoto18}.

\subsection{Mass -- (gas) metallicity relation}
The observed relationship between a galaxy's stellar mass and its metallicity
\citep[the mass$-$metallicity relation $-$ MZR, e.g.][]{Lequeux79,Tremonti04}
has been studied by many groups and using various methods to estimate metallicity.
The commonly employed set of methods make use of the optical emission lines coming from H II$-$regions
 to measure $Z_{O/H}$.
\\
The most direct approach is to measure auroral lines, which allow to estimate the electron temperature
of the gas within the region, that in turn is a strong function of metallicity 
\citep[so called direct method, e.g.][]{Stasinska05,AndrewsMartini13,Ly16}.
However, auroral lines are typically weak,
 especially at high metallicities and cannot be used as tracers at $Z_{O/H} \gtrsim Z_{O/H \odot}$.
Furthermore, direct method based $Z_{O/H}$ can be underestimated by even 0.4 dex in high-$Z_{O/H}$
environments if temperature gradients or fluctuations are present within the H II-region \citep[e.g.][]{Stasinska05}.
\\
To overcome those issues, several calibrations that allow to
translate ratios between the fluxes of strong emission lines into metallicity  have been
developed.
Those calibrations include empirical methods based on measurement of the electron temperature of the gas
\citep[e.g.][]{PettiniPagel04,Pilyugin05},
theoretical methods that rely on photoionization models 
\citep[e.g.][]{McGaugh91,KewleyDopita02,KobulnickyKewley04,Tremonti04}, 
or combination of the two \citep[e.g.][]{Denicolo02}.
The calibration dependent methods (referred to as strong line methods) lead to large differences
between the measured $Z_{O/H}$, with the offsets of even $\lesssim$0.7 dex \citep[][]{KewleyEllison08}.
The direct method and empirical calibrations typically lead to 2$-$3 times lower estimates than the theoretical ones, 
with combined methods falling in between. Those differences lead to different shapes and normalizations of the final MZR
\citep[][]{KewleyEllison08,MaiolinoMannucci18}.\\
Several studies used metal recombination lines \citep[][]{Esteban02,Bresolin07} that are weakly dependent on electron
temperature and insensitive to temperature fluctuations 
and may serve as an independent (next to emission line methods) indicator of metallicity.
However, observations of those faint lines are even more challenging than in the case of auroral lines.
Recombination line studies found metallicities consistent with those indicated by the 
calibrations based on photoionization models.
\\
Alternatively, the metallicity of the star formation can be estimated with the help of young, massive stars.
At the evolutionary timescale needed to form supergiants (a few 10 Myr) their host galaxy's interstellar medium
experiences very little chemical evolution and hence those stars are perfect candidates for that purpose.
Spectra of both blue and red supergiants were used to measure metallicities in galaxies even beyond the Local Group
\citep[e.g.][]{Kudritzki12,Kudritzki13,Kudritzki16,Lardo15,Bresolin16,Davies15,Davies17}.
Another approach to infer metallicity of the star formation using stars is to look at integrated light
spectra of young massive clusters with ages $<100 \rm \ Myr$ and typical masses $\sim10^{4} \Msun$
\citep[e.g.][]{Gazak14,Hernandez17,Hernandez18a}, which are mostly found in highly star forming galaxies.
Both approaches lead to metallicity estimates consistent with the direct method measurements
at low metallicities and O3N2/N2 empirical calibrations as found by \citet{PettiniPagel04} at high $Z_{O/H}$, 
being $\sim$0.4 dex lower than the 
metallicity estimates from calibrations based on photoionization models
\citep[e.g.][]{Davies17,Hernandez18a}.
\\
The source of these discrepancies between different methods is presently not clear, 
which makes it very difficult to set an absolute metallicity.
\\ \newline
\underline{Our choice:} 
Since the calibration leading to the correct estimation of the metallicity is not known,
we use the mass-metallicity relations based on 4 commonly used calibrations from the studies by:
\citet{Maiolino08} and refined by \citet{Mannucci09} (M09), \citet{Tremonti04} (T04), 
\citet{KobulnickyKewley04} (KK04) and \citet{PettiniPagel04} O3N2 calibration (PP04).\\
These relations cover the range of possible slopes and normalizations
for gas MZRs obtained with different methods to measure metallicity
\citep[e.g. Fig. 15 in][]{MaiolinoMannucci18}. See Sec. \ref{sec: method: MZR} for the details.

\subsubsection{Evolution with redshift}\label{sec: MZR: evolution}

In course of the history of the Universe next generations of stars form and evolve, 
gradually enriching the surrounding medium with metals. 
A fraction of this metal rich material can be lost from galaxies e.g. due to feedback 
from supernovae explosions or AGN activity or diluted by the inflowing metal$-$poor material.
Still, one would expect to see the imprint of this general enrichment as some form of 
the evolution in MZR with redshift.
The metallicity of distant galaxies can be estimated almost exclusively with
the strong emission line measurements.
Different emission line diagnostics are used at different redshifts, 
which may introduce artificial evolutionary trends or mask the true evolution of the relation.
Furthermore, those diagnostics are calibrated based on observations of local galaxies,
and it is not clear whether the conditions in which those lines are formed
do not change with $z$
\citep[e.g.][]{Kewley15}.
Reassuringly, \citet{Brinchmann08} argue that these effects do not strongly affect the 
abundance estimates from nebular lines and \citet{Patricio18} 
show that the commonly used diagnostics can be reliably applied up to z$\sim$2.
\\
Nonetheless, all of those issues make the study of the MZR evolution with redshift extremely difficult.
\citet{Moustakas11} study the MZR between z=0.05 and z=0.75 for a large sample of star forming galaxies
 with $M_{*}>10^{9.5} \Msun$. They use three different strong line theoretical calibrations
to estimate metallicities (McGaugh et al. (1991) et al.; Kobulnicky \& Kewley (2004) and Tremonti et al. (2004)).
Within the considered mass and redshift range they find no evidence for a mass dependent evolution 
and see a clear indication of decrease in metallicity with $z$ for all three calibrations.
However, the inferred rate of metallicity evolution is calibration dependent, with the slowest rate 
($- 0.16 \rm \ dex/z$) revealed when the \citet{KobulnickyKewley04} calibration is used, which also leads to the least steep MZR.
The other two calibrations lead to similar decrease rates ($\sim$ $- 0.26 \rm \ dex/z$).
\citet{Ly16} used the direct method to study the evolution of MZR at lower stellar masses
up to z$\sim$1 and also concluded that at a fixed stellar mass
within that redshift range metallicity decreases by around 0.25 dex when compared with the 
z$\sim$0 direct method based relation by \citet{AndrewsMartini13}
and the shape of the relation at z$\approx$0.5$-$1 is consistent with that found at z$\sim$0.
On the other hand, the results obtained by \citet{Maiolino08} and \citet{Mannucci09} who studied metallicities
of active galaxies at $z\sim3$ combined with z$\sim$2 measurements from \citet{Erb06}
indicate that MZR evolves in a mass dependent way, at least in the high-mass part ($M_{\ast}\gtrsim 10^{9}\Msun$) 
probed in those studies. They find that the characteristic mass at which the MZR flattens increases with $z$ 
and the overall evolution is stronger at lower masses.
A similar change in shape of the MZR was found by \citet{Zahid13}. Those authors argue that the mass independent
evolution reported by \citet{Moustakas11} may be caused by the differences in sample selection and the
fact that their sample was limited to considerably higher mass galaxies than in \citet{Zahid13}.
\\
The MZR can be studied with the emission lines up to z$\sim$3.5,
where the optical lines used to measure metallicity move out of the near IR bands
accessible to current ground-based spectrographs. 
Different tracers and methods need to be used beyond that redshift.
At cosmological distances metallicity can be estimated through absorption, 
e.g. using Damped Lyman Alpha systems \citep[e.g.][]{Wolfe86}. 
However, it is not clear how the metallicity measured
that way relates to the metallicities of star forming galaxies and how the two methods should 
be compared \citep[see sec. 3.6 in][]{MaiolinoMannucci18}.
\citet{Laskar11} used absorption lines in the interstellar medium of long gamma ray burst
host galaxies to infer the MZR up to z$\sim$5 and concluded that the
relation continues to evolve at z$>$3, although it is not clear whether the long GRB host galaxies
provide unbiased sample of high redshift star forming galaxies.
\\
Alternatively, one can use the information from the rest-frame UV spectra
that can be observed in the optical range for $z\gtrsim 2$ 
\citep[e.g.][note that those methods probe the metallicities of stellar populations rather than gas]{Rix04,Faisst16,Steidel16}.
\citet{Faisst16} validated the correlation between the metallicity and the equivalent width of
absorption features in the rest-frame UV 
 at z$\sim$2$-$3 and assuming that it also holds at higher $z$,
 applied that relation to probe the MZR at z$\sim$5.
 They find a very weak correlation between the stellar mass and UV based metallicity at z$\sim$5
within the probed mass/metallicity range, but their results come with large uncertainties
\citep[see discussion in Sec. 6 in][]{Faisst16}.
If those findings can be directly confronted with the gas-phase MZR from \citet{Mannucci09},
they show no evidence for evolution of the MZR since z$\sim$3.5.
However, the authors stress that the uncertainties involved in the measurements of various parameters 
(in particular metallicity and stellar mass) need to be reduced to allow for any firm conclusions.
\\ \newline
\underline{Our choice:} 
At z$\leq$3.5 we interpolate between the literature results.
The evolution of MZR is unconstrained at higher $z$. 
We assume that the normalization continues to decrease, while the shape remains the same as at z$\sim$3.5.
The rate of evolution depends on the metallicity calibration.\\
See Fig. \ref{fig: MZR_z} and Sec. \ref{sec: method: MZR} for the details.

\subsubsection{Scatter in the MZR} 

Metallicity as described by the MZR represents the average metal content of a galaxy
of a certain stellar mass and at a certain redshift.
\citet{Tremonti04} find that for a given stellar mass,
there is an intrinsic scatter of $\sigma\sim$0.1 dex (containing 68\% of the metallicity distribution)
around the best-fit mass $-$ metallicity relation. \citet{KewleyEllison08} find a similar scatter
of 0.08 $-$ 0.13 dex, depending on the metallicity calibration.\\
There is some evidence that the scatter may increase towards lower masses
\citep{Zahid14, Ly16}.
It is presently not clear whether the scatter in MZR evolves with redshift.
Determination of the intrinsic scatter is limited by the accuracy of the estimation
of observational uncertainties, which is more challenging in high redshift studies.
However, the analysis performed by \citet{Zahid14} suggests that there is no significant
evolution in the magnitude of the MZR scatter up z$\sim$0.8.
\\ \newline
\underline{Our choice:} 
For a given stellar mass and redshift we assume that there is a normally distributed scatter around
the mean gas metallicity given by the MZR, with the dispersion $\sigma_{0}$=0.1 dex 
(the intrinsic scatter in the MZR). 
We allow $\sigma_{0}$ to increase linearly with decreasing mass at M$_{*} < 10^{9.5} \Msun$.

\subsubsection{Distribution of metallicity within galaxies}\label{sec: MZR: gradients}

The average metallicity of a galaxy roughly corresponds to the metal content
that would be measured in this galaxy at the distance of $\sim \rm r_{e}$ 
one effective radius from its center \citep{KewleyEllison08}. 
Due to metallicity gradients present within galaxies there is a certain
range of metallicities with which stars can form inside their host.
The azimuthal metallicity variations are typically negligible \citep[e.g.][]{Sanchez-Menguiano17}.
However, typical radial abundance gradients in the local disk galaxies are around $-0.1$ dex/$r_{e}$ 
($\sim -0.03$ dex/kpc) \citep[e.g.][]{Sanchez14,Sanchez-Menguiano16}. 
\\
The detailed studies of the distribution of metallicities of HII regions within
z$\sim$0 disk galaxies carried out by \citet{Sanchez14}
and \citet{Sanchez-Menguiano16} with the CALIFA survey
show that the distribution of metallicities at which the star
formation proceeds is roughly symmetric with respect to the average
value found at $\sim \rm r_{e}$. 
However, the exact range of metallicities of the HII regions
and the metallicity gradient depend on the assumed
metallicity calibration \citep[see e.g. table 1][]{Sanchez-Menguiano16}
\footnote{
For instance, 75\% contours of the density distribution
of HII regions shown in Fig. 9 in \citet{Sanchez14} span the range of $\Delta_{Z_{O/H}}\sim$0.34 dex
in metallicity when the \citet{PettiniPagel04} O3N2 calibration is used.
\citet{Sanchez-Menguiano16} found a narrower range of $\Delta_{Z_{O/H}}\sim$0.22 dex.
However, the metallicity calibration used in this study \citep[O3N2 calibration by][]{Marino13} 
leads to shallower metallicity gradients.
In both cases the range is roughly symmetric with respect to the metallicity corresponding to the
value measured at r$_{e}$, with the HII regions in the inner parts of galaxies having metallicities
ranging from $Z_{O/H}$(r$_{e}$) to $Z_{O/H}$(r$_{e}$)+$\Delta_{Z_{O/H}}$/2 and the outer parts from 
$Z_{O/H}$(r$_{e}$) to $Z_{O/H}$(r$_{e}$)-$\Delta_{Z_{O/H}}$/2. 
}.
\\ \newline

\underline{Our choice:} 
We assume that the distribution of metallicities at which the star formation proceeds
within galaxies can be represented by a normal distribution with dispersion $\sigma_{\nabla Z_{O/H}}$=0.14 dex.
This particular choice of $\sigma_{\nabla Z_{O/H}}$ corresponds to half of the average range of metallicities
 of HII regions$^{2}$ found by \citet{Sanchez14} and \citet{Sanchez-Menguiano16}, 
 averaged between the two studies.\\
 See appendix \ref{app: secondary assumptions} for additional discussion.

\subsection{Star formation -- mass relation}

The stellar mass and star formation rate of star forming galaxies are strongly correlated,
giving rise to the star formation -- mass relation 
\citep[SFMR; e.g.][]{Brinchmann04, Salim07, Speagle14, Tomczak16, Boogaard18} 
also called the star forming main sequence.
\\
SFRs are measured based on the luminosities of galaxies
observed in certain bands that correlate with the recent ($\lesssim$ 100 Myr)
star formation activity.
Those luminosities need to by corrected for dust attenuation, 
which is one of the main sources of uncertainty in SFR estimation.
Furthermore, conversion from luminosity to SFR is sensitive
to the assumed initial mass function (IMF) and metallicity.
The most common tracers of SFR are UV luminosity (usually at $\sim$1500 -- 2800 $\AA$),
certain recombination lines e.g. H$\alpha$ and IR continuum emission between $\sim$ 3 $-$ 1100 $\mu$m.
Different indicators are sensitive to the star formation on different timescales:
H$\alpha$ line typically probes short $\sim$10 Myr timescales
while UV and IR provide information on SFR averaged over longer $\sim$100 Myr timescales
\citep[see e.g.][for a detailed discussion of different SFR tracers]{KennicuttEvans12,MadauDickinson14}.
\\
The star formation main sequence at a certain redshift is commonly described as a power law relation
connecting the galaxy stellar mass M$_{\ast}$ and SFR
\begin{equation}\label{eq: SFMR}
 \rm log\left(SFR/\Msun yr^{-1} \right) = a \times log\left(M_{\ast}/\Msun \right) + b
\end{equation}
with coefficients $a$ and $b$ describing slope and normalization of the relation respectively.
The typically found low-mass end slope values fall between 0.75 and 1 \citep[see fig. 10 in][]{Boogaard18}.
Some studies find departure from the single power-law given by eq. \ref{eq: SFMR},
identifying a flattening at the high mass end 
\citep[i.e. galaxies with M$_{\ast} \lesssim 10^{10} \Msun$ follow a steeper relation
than their more massive counterparts, e.g.][]{Whitaker14,Lee15,RenziniPeng15,Schreiber15,Tomczak16},
while the other find no evidence for such a turnover \citep[e.g.][]{Speagle14,Pearson18}.
Hence, the presence and the amount of flattening in the SFMR is currently not clear.
\citet{Johnston15} suggest that the presence of the turnover may depend on the method
used to filter out the quiescent galaxies from the sample used to construct the SFMR.
\\ \newline
\underline{Our choice:}
As a base relation we adopt the recent result obtained by \citet{Boogaard18},
who focused on the lower mass part of the SFMR where it can be described as a single power law.
Since the presence (and the degree) of the flattening at the high mass end of the relation is debatable,
we explore three variations of its shape: single power law at all masses (no flattening),
broken power law (moderate flattening) and almost constant SFR at high masses (sharp flattening).
\\
The details can be found in Sec. \ref{sec: method: SFMR-z}.

\subsubsection{Evolution with redshift}\label{sec: SFMR-z}

\citet{Speagle14} converted results from 25 studies probing M $\gtrsim 10^{10} \Msun$ and 
reaching up to redshift z$\sim$5 to a common set of calibrations and
concluded that the high mass end slope of the SFMR  shows a mild
redshift evolution (see Fig. 8 therein). 
To our knowledge there is no similar analysis constraining the lower part of the SFMR.\\
Regardless of the precise form of the SFMR, its normalization is known to evolve with redshift.
This evolution is commonly parametrized by adding a factor 
$c\times \rm log(1+z)$ to eq. \ref{eq: SFMR} with power law exponent $c$.
Values reported in the literature range from c$\sim$1.8 up to c$\sim$4 at redshifts z$\lesssim$2
\citep[][]{Karim11, Speagle14,Whitaker14,Ilbert15,Lee15,Tasca15,Schreiber15,Tomczak16,Boogaard18}.
Some studies find evidence for higher values of $c$ at high mass part of the SFMR
\citep[$\gtrsim 10^{10}\Msun$ e.g.][]{Whitaker14,Ilbert15} than at the low mass part.
Combining different datasets, \citet{Speagle14} find c$\sim$2.8.
At z$\gtrsim$2 the SFMR normalization shows little to no evolution
\citep[c$\lesssim$1  e.g.][]{Gonzalez14,Tasca15,Santini17,Pearson18}
in tension with theoretical predictions \citep[e.g.][]{Weinmann11}, based on which
one would expect the SFMR to decrease monotonically with cosmic time with c$\sim$2.2-2.5.
\\ \newline
\underline{Our choice:}
We assume that the low mass end slope of the SFMR does not evolve with redshift.
The evolution of the high mass part depends on the variation (see Sec. \ref{sec: method: SFMR-z}).
The normalization increases with redshift as $c\times \rm log(1+z)$
 with $c$=2.8 at z$\leq$1.8 following \citet{Speagle14}. 
 This value falls roughly in the middle of the range of values found in the literature.
 At higher redshifts we use $c$=1 to reproduce the observed flattening
 in the redshift evolution of SFMR normalization.
 The redshift at which we change the value of $c$ corresponds to the redshift
 of the peak in the cosmic star formation history \citep[e.g.][]{MadauDickinson14,MadauFragos17,Fermi18}.
 The redshift evolution of our SFMR is shown in Figure \ref{fig: SFRM_z}.

\subsubsection{Scatter in the relation}
Similarly to the MZR, there is 'intrinsic' scatter ($\sigma_{\rm SFR}$) in the SFMR.
Its determination requires disentangling the
measurement error, redshift evolution within the sampled range
and the intrinsic scatter and proves challenging.
\citet{Speagle14} found $\sigma_{\rm SFR}$=0.2 dex and
a similar value was recently obtained by \citet{Pearson18}.
Those values are on the lower side of estimates present in the literature
which report the SFMR scatter about $\sim$0.3 $-$ 0.4 dex 
\citep[e.g.][]{Salim07,Whitaker12,RenziniPeng15,MattheeSchaye19},
while \citet{Kurczynski16} and \citet{Boogaard18} found $\sigma_{\rm SFR}$ as high as 0.44 dex.
As argumented by \citet{Boogaard18}, those differences may be partially attributed to different SFR timescales
probed by different indicators.
They also point out that using the same data to derive SFR and stellar masses induces
correlation between the two which might artificially decrease the scatter found in the studies that do so.
\citet{Salim07} suggest that the scatter may increase towards lower stellar masses, 
but no clear trend neither with mass nor with redshift was found in other observational studies 
\citep[e.g.][]{Whitaker12,Schreiber15}.
\\ \newline
\underline{Our choice:}
For any $M_{*}$ and $z$ we assume that the SFR is normally distributed around the mean given
by the SFMR with the dispersion $\sigma_{\rm SFR}$=0.3 dex.

\subsection{Fundamental metallicity relation}

\citet{Ellison08} suggested that there is a more general relation connecting
all three quantities characterizing galaxies discussed above: 
stellar mass, gas-phase metallicity and star formation rate, 
further investigated by \citet{Mannucci10} and called fundamental metallicity relation (FMR).
The relation is such that galaxies of the same stellar mass showing higher than average SFR 
also have lower metallicities.
A similar correlation was found in a number of studies
\citep[e.g.][]{AndrewsMartini13,Lara-Lopez13,Salim14,Zahid14,Yabe15}.
The correlation weakens/ceases at very high stellar masses $M_{\ast}\gtrsim 10^{11} \Msun$.
\citet{Mannucci10} found no evolution in the FMR up to the redshift of $z\sim2.5$, 
i.e galaxies at those redshifts
are found on the same $M_{*}-$metallicity$-$SFR 3D plane as the local galaxies, contrary to \citet{Zahid14}
who found the redshift dependence of FMR.
The exact form of the observationally inferred FMR depends on the method used to select galaxies, 
measure metallicity, SFR and stellar masses.
Variations in $\alpha$-enhancement may also influence the observational estimates of FMR \citep{MattheeSchaye18}.
\\ \newline
\underline{Our choice:}
The existence of the FMR, regardless of its precise functional form, means that
we cannot choose SFR and metallicity of a galaxy of a certain mass independently.
To account for the anti-correlation in SFR and metallicity for a given $M_{\ast}$,
we assume the scatter in both relations is anti-correlated
\footnote{For instance, if a galaxy's SFR (chosen from a
normal distribution with $\sigma_{\rm SFR}$=0.3 dex and mean $\mu_{\rm SFR}$ given by the SFMR) 
is higher than indicated by the mean then the metallicity assigned to that galaxy will be lower
than the mean $\mu_{\rm MZR}$ resulting from the MZR
(and vice versa; e.g. if $\rm SFR(M_{\ast},z)=\mu_{\rm SFR}+\frac{\sigma_{\rm SFR}}{3}$ then
$\rm Z(M_{\ast},z)=\mu_{\rm MZR}-\frac{\sigma_{\rm 0}}{3}$).
}

\subsection{Initial stellar mass function}\label{sec: IMF}

The IMF influences many of the observable properties of stellar populations and galaxies (e.g. SFR,M$_{*}$)
and hence also the empirical scaling relations.
In this paper we assume a non-evolving, universal IMF.
This is a common assumption and it is already introduced within the observational
studies whose results we use in our work.
\\
However, the IMF has been theoretically predicted to vary with star-forming conditions
in such a way that in metal poor environments and/or regions with warmer gas the IMF becomes more top-heavy
\citep[e.g. see review by][and references therein]{Kroupa13}.
Studies based on the resolved stellar populations in Local Group galaxies
reveal no clear evidence for systematic trends in 
the IMF with metallicity (e.g. \citealt{Bastian10, Kroupa13, Offner14}, but see \citealt{Hopkins18}).
Some evidence for such variations has been found in studies of globular clusters
or ultra compact dwarf galaxies and within starburst galaxies \citep[e.g.][]{Kroupa13,Schneider18,Zhang18}.
Furthermore, recent studies found radial IMF variations within massive early type galaxies, 
with the IMF generally found to be more bottom-heavy in their centers 
(e.g. \citealt{Martin-Navarro15,LaBarbera16,OldhamAuger18}, but see \citealt{Smith14,Smith15}).
While there is growing evidence for no general universality of the IMF,
it is presently not clear how the IMF varies with environment or redshift 
(and to what extent those variations can arise due to the differences in methodology).
Self-consistent modeling of the variable IMF is challenging
and the proposed models struggle to reproduce all observational constraint simultaneously
(e.g. \citealt{Barber18,Barber19}, but see \citealt{Weidner13,Jerabkova18,Yan17})
\\
We note that significant variations of the IMF (e.g. with metallicity), if present,
could have a strong and non-straightforward effect on our results.
However, with the current state of knowledge a meaningful assessment of this effect 
is a complex (if not impossible) task and is beyond the scope of this study.
\\ \newline

\underline{Our choice:}
We assume a \citet{Kroupa01} IMF with the mass limits 0.1 $-$ 100 $\Msun$ (independent of the environment or redshift).
We correct for the differences between the IMFs adopted in various studies as described in the appendix
\ref{app: method: IMF}.

\section{Method}\label{sec:method}

\begin{figure*}
\centering
\includegraphics[scale=0.46]{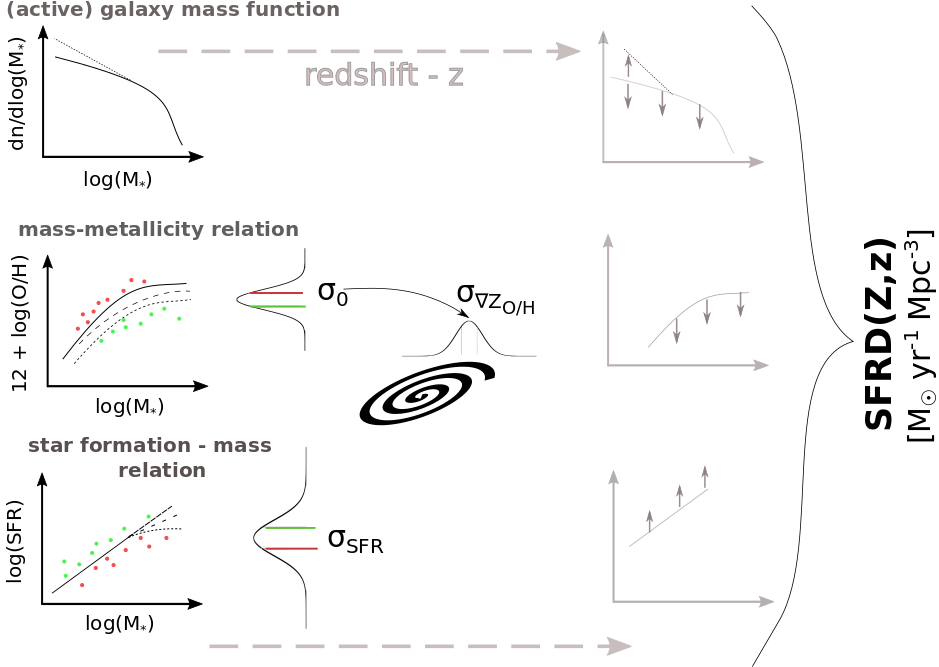}
\caption{ 
Schematic representation of our method.
The observed stellar mass function of star forming galaxies gives 
the number density of objects of different masses.
The mass $-$ metallicity relation allows us to assign metallicities to galaxies of different stellar masses.
The star formation - mass relation allows us to specify the contribution of galaxies of different
masses (metallicities) to the total star formation rate density at a certain redshift.\newline
We vary the assumptions about those relations 
(low mass end slope of the GSMF, normalization and shape of the MZR, high mass end of the SFMR)
to cover the range of possibilities present in the literature.
We account for the intrinsic scatter present in the relations ($\sigma_{0}$, $\sigma_{SFR}$)
and the observed anti-correlation between the SFR and metallicity.
On top of that we introduce scatter in metallicity to account for the internal distribution
of metallicities in the star forming gas within galaxies ($\sigma_{\nabla Z_{O/H}}$).
All relations evolve with redshift. \newline
Combining all relations we obtain the distribution of the cosmic star formation
rate density over metallicities and redshifts SFRD(Z,z).
}
\label{fig: method}
\end{figure*}

To construct our model and calculate the distribution of star formation rate density across 
different metallicities and redshifts,
we combine the observationally inferred galaxy stellar mass function (GSMF) of star forming galaxies
with observational relations connecting stellar masses of galaxies with their gas-phase metallicity ($Z_{O/H}$)
and star formation rate (SFR). Our method is schematically summarized in Figure \ref{fig: method}.\\
The procedure can be summarized as follows:
\begin{itemize}
 \item 
  We consider galaxies with stellar masses between $10^{6} - 10^{12} \Msun$
  and divide this mass range into $N_{gal}$ mass bins $\Delta M_{j}$ equally spaced in logarithm.
  Within each redshift/time bin we calculate the number density of galaxies falling within
  each mass bin $n_{\Delta M;j}$, using the galaxy stellar mass function of star forming galaxies.
  This mass density is assumed to be constant within the redshift/time bin.
 \item From each mass bin we randomly choose $N_{sampl}$ stellar masses. 
  Each of those masses $M_{i}$ is associated with a certain number density
  of galaxies $n_{M_{i}}=\frac{n_{\Delta M;j}}{N_{sampl}}$
 \item 
  Each mass $M_{i}$ is then assigned a certain star formation rate $SFR_{i}$, 
  drawn from a SFMR for a given redshift and taking into account the scatter in the relation. 
  This SFR is assumed to be constant within the redshift/time bin.
 \item 
 For each mass $M_{i}$ we draw the corresponding gas metallicity $Z_{O/H \ i}$ from the MZR 
  at a given redshift, taking into account
  the scatter in the relation and correlation between the SFR and metallicity.
   Again, $Z_{O/H \ i}$ is assumed to be constant within the redshift/time bin.
  \item 
  We then calculate the amount of mass formed in stars within the 
  time/redshift bin per unit of comoving volume in galaxies of different masses
   and metallicities
 and sum the contributions from galaxies that fall within the same metallicity bin.
\end{itemize}
We use different variations of the assumptions to estimate the uncertainties of our results.
In the following subsections we provide the details of the construction of our model and
describe how the different observational results introduced in Sec. \ref{sec: section2} are combined.

\subsection{ Galaxy stellar mass function}\label{sec: method: GSMF}	

\begin{figure}
\centering
\includegraphics[scale=0.4]{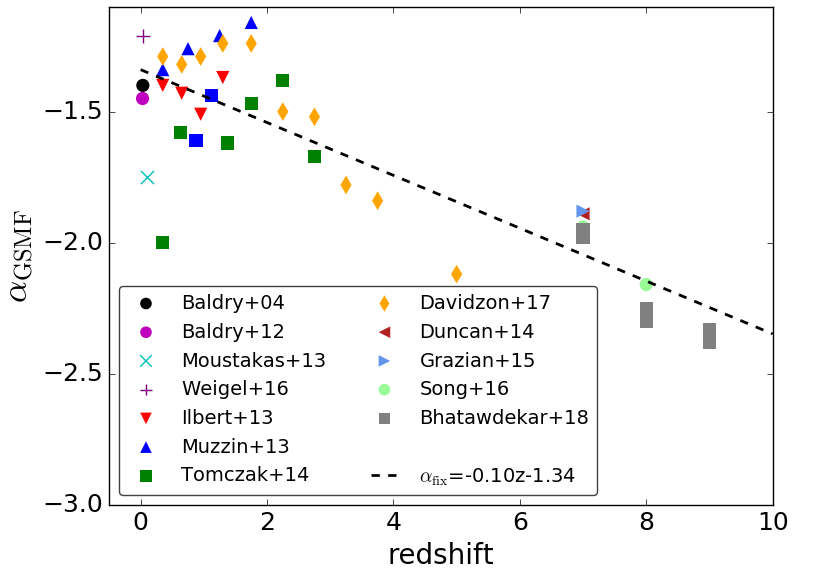}
\caption{ 
Low mass end slope ($\alpha_{\rm GSMF}$) values at different redshifts resulting 
from best (double) Schechter function fits to the galaxy stellar mass function of
star forming galaxies ($z<6$) or GSMF of all galaxies ($z>$6) as found by different authors
(see Tab. \ref{tab: GSMF}; the fixed $\alpha_{\rm GSMF}$ cases from Ilbert et al. (2013)
and Muzzin et al. (2013) were excluded from the fit).
The dashed line shows the linear fit used to describe the evolution of $\alpha_{\rm fix}(z)$
in those variations of our model in which we allow this slope to vary with redshift.
}
\label{fig: alpha_GSMF_vs_z}
\end{figure}

\begin{figure}
\centering
\includegraphics[scale=0.35]{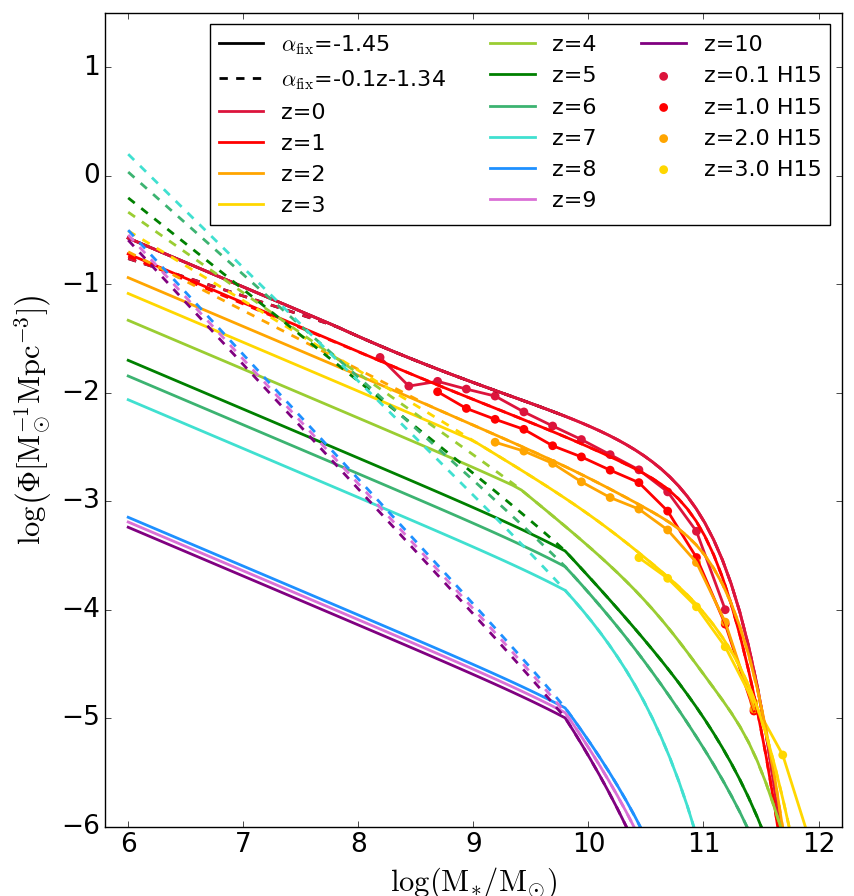}
\caption{ 
Galaxy stellar mass function of star forming galaxies used in our model, shown at different redshifts $z$.
The solid lines show GSMF with the low mass end slope fixed to $\alpha_{\rm fix}$=$-$1.45,
while the dashed lines show the case when the slope is allowed to evolve (steepen) with redshift.
The observation-based GSMF of blue galaxies used by \citet{Henriques15} to constrain their semi-analytic
model is shown with the dot-connected lines at z=0.1, 1, 2 and 3 for comparison.
}
\label{fig: GSMF_vs_z}
\end{figure}

\begin{figure*}
\centering
\includegraphics[scale=0.37]{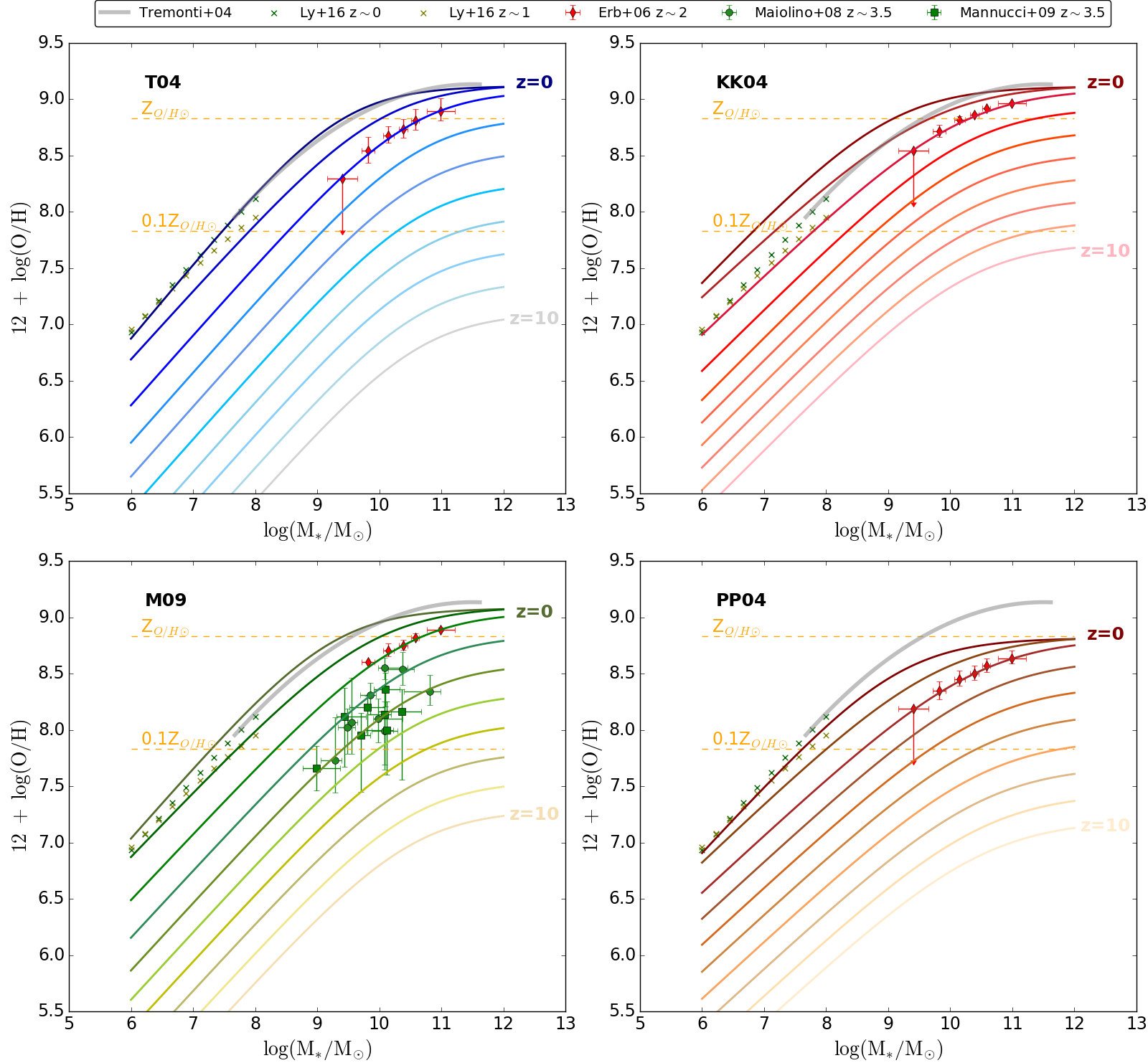}
\caption{ 
 The four versions of the mass $-$ metallicity relation (MZR) used in this study.
 The relations were obtained using the results from \citet{Mannucci09} (M09) and
 converting them to different metallicity calibrations
 (T04 $-$ \citet{Tremonti04}, KK04 $-$ \citet{KobulnickyKewley04}, PP04 $-$ \citet{PettiniPagel04} O3N2; see text).
 Each panel shows ten lines, corresponding to the MZR at different redshifts 
 (z=0 $-$ top lines, z=10 $-$ bottom lines, $\Delta$z=1 spacing between the lines).
 The orange dashed horizontal lines mark the solar and 10\% solar metallicity,
 assuming $Z_{O/H \odot}$=8.83 \citep{AndersGrevesse89}.
 For the reference, we plot the local MZR from \citet{Tremonti04} at high stellar masses (thick gray line) and 
 the MZR from \citet{Ly16} at low stellar masses at z$\sim$0 and z$\sim$1 (dark and light green crosses respectively) 
 in each panel. We also plot the data points from \citet{Erb06} at z$\sim$2 (red points), converted to different
 metallicity calibrations where necessary. In the bottom left panel we additionally plot the data from
 \citet{Maiolino08} and \citet{Mannucci09} at z=3$-$4. We correct for different IMF where necessary.
}
\label{fig: MZR_z}
\end{figure*}

We use the published best-fit parameters to the Schechter (or double Schechter) function describing
GSMF of star forming galaxies in different redshift bins at z$\lesssim$6. 
The studies used in the analysis are listed in Table \ref{tab: GSMF}.
The last column of Table \ref{tab: GSMF} gives the limits of the redshift bins 
in which the GSMF was estimated in each of those studies.
We choose several values of redshift $z_{\rm avg}$
that overlap with the redshift bins in at least three of those studies
\footnote{
with the exception of z=5 where we use only one study by \citet{Davidzon17}, 
as this is the only result constraining GSMF of active galaxies at z$>$4.
}. 
Those values are listed in the first column of Table \ref{tab: GSMF}.
\\
To obtain the number density of star forming galaxies of a certain mass
and at a certain redshift we:
\\
(i) calculate the number density for that mass at $z_{\rm avg}$ redshifts
as the average from the number densities calculated using the Schechter function fits
referenced in Tab. \ref{tab: GSMF}
\\
(ii) interpolate between the values found at different redshifts $z_{\rm avg}$
\\
We treat the low and high mass part of the GSMF separately, 
calculating the number density of galaxies using the full Schechter fits as
described above at $M_{\ast}>M_{\rm GSMF}$ and assuming that at lower masses
the GSMF of star forming galaxies is described as a single power law with slope $\alpha_{\rm fix}$.
We allow for two variations: 
\begin{itemize}
 \item $\alpha_{\rm fix}$=$-$1.45 at all redshifts
 \item $\alpha_{\rm fix}$=$\alpha_{\rm fix}(z)$ steepening with 
      redshift according to the linear fit shown in Fig. \ref{fig: alpha_GSMF_vs_z} 
      up to z=8 and fix $\alpha_{\rm fix}$=$-$2.14 at z$>$8 \footnote{
      The high redshift study by \citet{Bhatawdekar18} shows little to no evolution of the GSMF at z$\gtrsim$8
      }.
\end{itemize}
We assume that the mass separating the low and high mass part of the GSMF increases with redshift as
log$\left( M_{\rm GSMF}/\Msun \right) = 7.8 + 0.4 z$
\footnote{
This roughly corresponds to the mass completeness limit as a function of redshift from \citet{Tomczak14}
(see fig. 2 therein). This limit in other $z<5$ studies is typically higher, thus we allow for some
extrapolation of the fitted Schechter function beyond that limit before fixing the slope.
}
up to $z$=5 and equals log$\left( M_{\rm GSMF}/\Msun \right)$=9.8 at
higher redshifts.
\\
At $z<$0.05 we use the estimate obtained for $z$=0.05.
To our knowledge there are no measurements of GSMF for the star forming population of galaxies at $z>$6.
To extend our analysis to higher redshifts and constrain the GSMF evolution, 
we include the recent results obtained for the total sample of galaxies
at $z\sim$7, $z\sim$8 and z$\sim$9 \citep{Duncan14,Grazian15,Song16,Bhatawdekar18}. 
This is justified as at such high redshifts the fraction of passive galaxies is expected to be very low
\citep[e.g.][]{Muzzin13,Henriques15,Davidzon17}.
At $z>$9 we assume that the normalization of the GSMF continues to mildly decrease
at the rate found between z=8 and z=9.
The resulting GSMF is shown in Fig. \ref{fig: GSMF_vs_z}.

 \begin{table}
\centering
\small
\caption{
The first column gives the redshift at which we average over the relations from
studies listed in the second column in our model.
The third column gives the IMF used in each study.
The fourth column provides the slope of the low mass end of the star forming GSMF
as fitted by the authors and the last one ($z_{bin}$) provides the redshift range of galaxies
used by the authors to construct a GSMF of active galaxies.
At $z$=0.95 we combine two redshift bins from \citet{Tomczak14} and at $z$=3.5 two redshift
bins from \citet{Davidzon17}, since those redshifts fall at the edge of their bins.
\newline
$^{a}$: using parameters fitted by \citet{Tomczak14} \newline
$^{b}$: the steeper slope from the fit to a double Schechter function \newline
$^{c}$: the value was fixed during the fitting procedure \newline
$^{d}$: GSMF describing the `total` sample of galaxies \newline
$^{e}$: fits in parentheses include galaxies flagged by \citet{Bhatawdekar18} as `point sources`
}
\begin{tabular}{c c c c c }
\hline
 z avg. & reference & IMF & $\alpha_{\rm GSMF}$ & z$_{bin}$ \\ \hline
  \hline
\multirow{4}{*}{0.05} & \citet{Baldry04}  & K01  & $-$1.4$^{b}$ & 0.01$-$0.08\\ 
		      & \citet{Baldry12}  & Ch03 & $-$1.45 & 0.02$-$0.06\\ 
		      & \citet{Moustakas13}$^{a}$ & Ch03 & $-$1.75$^{b}$ & 0.01$-$0.2\\
		      & \citet{Weigel16}  & K01  & $-$1.21 & 0.02$-$0.06\\ \hline
\multirow{4}{*}{0.35} & \citet{Ilbert13}  & Ch03 & $-$1.4$^{b}$  & 0.2$-$0.5\\ 
		      & \citet{Muzzin13}  & K01  & $-$1.34       & 0.2$-$0.5\\ 
		      & \citet{Tomczak14} & Ch03 & $-$2$^{b}$    & 0.2$-$0.5\\
		      & \citet{Davidzon17}& Ch03 & $-$1.29$^{b}$ & 0.2$-$0.5\\ \hline	
\multirow{4}{*}{0.65} & \citet{Ilbert13}  & Ch03 & $-$1.43$^{b}$ & 0.5$-$0.8\\ 
		      & \citet{Muzzin13}  & K01  & $-$1.26       & 0.5$-$1\\ 
		      & \citet{Tomczak14} & Ch03 & $-$1.58$^{b}$ & 0.5$-$0.75\\
		      & \citet{Davidzon17}& Ch03 & $-$1.32$^{b}$ & 0.5$-$0.8\\ \hline
\multirow{4}{*}{0.95} & \citet{Ilbert13}  & Ch03 & $-$1.51$^{b}$ & 0.8$-$1.1\\ 
		      & \citet{Tomczak14} & Ch03 & $-$1.61$^{b}$ & 0.75$-$1\\ 
		      & \citet{Tomczak14} & Ch03 & $-$1.44$^{b}$ & 1$-$1.25\\
		      & \citet{Davidzon17}& Ch03 & $-$1.29$^{b}$ & 0.8$-$1.1\\ \hline
\multirow{4}{*}{1.3} & \citet{Ilbert13}  & Ch03 & $-$1.37$^{b}$ & 1.1$-$1.5\\ 
		      & \citet{Muzzin13}  & K01  & $-$1.21       & 1$-$1.5\\ 
		      & \citet{Tomczak14} & Ch03 & $-$1.62$^{b}$ & 1.25$-$1.5 \\
		      & \citet{Davidzon17}& Ch03 & $-$1.24$^{b}$ & 1.1$-$1.5 \\ \hline	
\multirow{4}{*}{1.75} & \citet{Ilbert13}  & Ch03 & $-$1.6$^{b,c}$ & 1.5$-$2\\ 
		      & \citet{Muzzin13}  & K01  & $-$1.16       & 1.5$-$2\\ 
		      & \citet{Tomczak14} & Ch03 & $-$1.47$^{b}$ & 1.5$-$2 \\
		      & \citet{Davidzon17}& Ch03 & $-$1.24$^{b}$ & 1.5$-$2 \\ \hline	
\multirow{4}{*}{2.25} & \citet{Ilbert13}  & Ch03 & $-$1.6$^{b,c}$& 2$-$2.5\\ 
		      & \citet{Muzzin13}  & K01  & $-$1.3$^{c}$  & 2$-$2.5\\ 
		      & \citet{Tomczak14} & Ch03 & $-$1.38$^{b}$ & 2$-$2.5 \\
		      & \citet{Davidzon17}& Ch03 & $-$1.5$^{b}$  & 2$-$2.5 \\ \hline
\multirow{4}{*}{2.75} & \citet{Ilbert13}  & Ch03 & $-$1.6$^{b,c}$& 2.5$-$3\\ 
		      & \citet{Muzzin13}  & K01  & $-$1.3$^{c}$       & 2.5$-$3\\ 
		      & \citet{Tomczak14} & Ch03 & $-$1.67$^{b}$ & 2.5$-$3 \\
		      & \citet{Davidzon17}& Ch03 & $-$1.52$^{b}$  & 2.5$-$3 \\ \hline
\multirow{4}{*}{3.5}  & \citet{Ilbert13}  & Ch03 & $-$1.6$^{b,c}$& 3$-$4\\ 
		      & \citet{Muzzin13}  & K01  & $-$1.3$^{c}$  & 3$-$4\\ 
		      & \citet{Davidzon17}& Ch03 & $-$1.78  & 3$-$3.5 \\
		      & \citet{Davidzon17}& Ch03 & $-$1.84  & 3.5$-$4 \\ \hline
\multirow{1}{*}{5}    & \citet{Davidzon17}& Ch03 & $-$2.12  & 4$-$6 \\ \hline \hline      
\multirow{4}{*}{7}    & \citet{Duncan14}$^{d}$  & Ch03 & $-$1.89 & 6.5$-$7.5\\ 
		      & \citet{Grazian15}$^{d}$  & Sal  & $-$1.88  & 6.5$-$7.5\\ 
		      & \citet{Song16}$^{d}$ & Sal & $-$1.94 & 6.5$-$7.5 \\
		      & \citet{Bhatawdekar18}$^{d,e}$& Ch03 & $-$1.98 ($-$1.95) & 6.5$-$7.5 \\ \hline
\multirow{2}{*}{8}    & \citet{Song16}$^{d}$ & Sal & $-$2.16 & 7.5$-$8.5 \\
		      & \citet{Bhatawdekar18}$^{d,e}$& Ch03 & $-$2.3 ($-$2.25) & 7.5$-$8.5 \\ \hline
		9     & \citet{Bhatawdekar18}$^{d,e}$& Ch03 & $-$2.38 ($-$2.33)& 8.5$-$9.5 \\ \hline
\end{tabular}
\label{tab: GSMF}
\end{table}
 
\subsection{Mass-(gas) metallicity relation} \label{sec: method: MZR}

We start from the mass metallicity relation fitted by \citet{Maiolino08} 
(in 3 redshift bins z$\sim$ 0.07, 0.7 and 2.2)
and refined by \citet{Mannucci09} (at z$\sim$3.5)
\footnote{
We convert the results from \citet{Maiolino08} to \citet{Chabrier03} IMF used by \citet{Mannucci09} 
by applying the correction suggested by these authors and then increase the
logarithm of the mass by 0.03 dex to convert these results to \citet{Kroupa01} IMF.
}
We use a different parametrization of the MZR than in those two studies, given by eq. \ref{eq: MZR}
and proposed by \citet{Moustakas11}.
\begin{equation}\label{eq: MZR}
 \rm 12 + log[O/H] = Z_{O/H \ asym} - log\left[ 1 + \left(\frac{M_{\ast}}{M_{TO}} \right)^{-\gamma} \right]
\end{equation}
This parametrization allows to avoid an artificial turn-off at the high mass part of the relation.
The parameters are: $\gamma$ describing the low--mass end slope,
$M_{\rm TO}$ -- the mass at which the relation begins to turn/flatten and $Z_{O/H \ asym}$ -- the
asymptotic metallicity of the high-mass end.
The relation given by \citet{Mannucci09} was refitted in each of the redshift bins using this 
parametrization.
We refer to this form of MZR as $M09$ throughout the paper.
The gas metallicities in \citet{Mannucci09} at 12+log[O/H]$>$8.35 dex were obtained with the
 Kewley \& Dopita (2002) strong line calibration.
We translate the base $M09$ relation at each redshift
to three other calibrations 
(\citet{Tremonti04} -- $T04$, \citet{KobulnickyKewley04} -- $KK04$ and \citet{PettiniPagel04} O3N2 -- $PP04$) 
applying the conversion method described in \citet{KewleyEllison08} at the 
high metallicity part (12+log[O/H]$>$8.35 dex) and fit the relation given by eq. \ref{eq: MZR}
to the converted MZRs.

The fitted parameters are summarized in Table \ref{tab: MZR}
\footnote{The slope of the $PP04$ MZR was fixed at z$>$2.2 to the value fitted for that redshift 
(could not be fitted with $M09$ data at metallicities higher than 8.35 alone)}.
\\
We interpolate between the redshift bins to obtain the MZR at $0 \leq \rm z \leq 3.5$ for each calibration.
We assume that at higher redshifts the relation evolves in a mass independent fashion, i.e. only the
normalization decreases. The rate of decrease in normalization (dZ$_{O/H}$/dz)
at z$>$3.5 is assumed to be the same as the decrease in metallicity at $\rm M_{\ast}=10^{11} \Msun$ between 
the two highest redshift bins (z=2.2 and z=3.5).
Hence, for a given calibration the MZR at z$>$3.5 is given by:
\begin{equation}\label{eq: MZR(z)}
 \rm \left(12 + log[O/H]\right)_{z>3.5} = \left(12 + log[O/H]\right)_{z=3.5} + dZ_{O/H}/dz \times (z-3.5)
\end{equation}
For a galaxy with mass $M_{*}$ at redshift $z$, we draw its average metallicity
from the normal distribution centered at the metallicity taken from the MZR
and with dispersion $\sigma_{0}$=0.1 dex at $M_{*}>10^{9.5} \Msun$ 
and $\sigma_{0}$=--0.04 log(M${*}$/$\Msun$)+0.48 dex at higher masses.
We add a normally distributed scatter around this value with $\sigma_{\nabla Z_{O/H}}$=0.14 dex
to account for the distribution of metallicities at which the stars are forming within their host galaxies.
\\

 \begin{table}
\centering
\small
\caption{
The parameters of the different versions of the mass $-$ metallicity relation used in this study.
The first column gives the redshift, the second (fifth) column provides the low-mass end slope of the relation
$\gamma$ , the third (sixth) column gives the logarithm of the turnover mass log$\left( \rm M_{TO} \right)$
and the fourth (seventh) column the asymptotic metallicity of the high mass end of the relation Z$_{O/H \ asym}$
(see eq. \ref{eq: MZR}).
dZ$_{O/H}$/dz gives the rate of evolution of the MZR normalization at redshifts $>3.5$ (see eq. \ref{eq: MZR(z)}).
}
\begin{tabular}{c c c c c c c}
\hline
 z & $\gamma$ & log$\left( \rm M_{TO} \right)$ & Z$_{O/H  asym}$ &  $\gamma$ & log$\left( \rm M_{TO} \right)$ & Z$_{O/H  asym}$  \\ \hline
  \hline
 &\multicolumn{3}{c}{ T04; dZ$_{O/H}$/dz=$-$0.29 dex } &   \multicolumn{3}{c}{ M09; dZ$_{O/H}$/dz=$-$0.26 dex }\\ \hline
 0   & 0.66 & 9.39 &  9.12  & 0.63 & 9.25 &  9.08 \\
 0.7 & 0.61 & 9.86 &  9.15  & 0.57 & 9.72 &  9.11 \\
 2.2 & 0.62 & 10.59 & 9.07  & 0.59 & 10.46 &  9.04 \\
 3.5 & 0.62 & 10.67 & 8.70  & 0.60 & 10.54 &  8.72 \\ \hline
 &\multicolumn{3}{c}{ KK04; dZ$_{O/H}$/dz=$-$0.20 dex } &  \multicolumn{3}{c}{ PP04; dZ$_{O/H}$/dz=$-$0.24 dex } \\ \hline
 0   & 0.57 & 9.03 &  9.12  & 0.60 & 9.19 &  8.81 \\
 0.7 & 0.51 & 9.49 &  9.14  & 0.53 & 9.67 &  8.85\\
 2.2 & 0.53 & 10.26 & 9.09  & 0.51 & 10.54 &  8.81  \\
 3.5 & 0.56 & 10.32 & 8.83  & 0.51 & 10.54 &  8.52 \\ \hline

\end{tabular}
\label{tab: MZR}
\end{table}

\subsection{Star formation - mass relation} \label{sec: method: SFMR-z}
 Following \citet{Boogaard18}, we describe the low mass part of the SFMR 
 as a single slope power law with $a$=0.83 (see eq. \ref{eq: SFMR}).
 This slope does not evolve with redshift.
 We explore three variations of the SFMR shape at high masses:

 \begin{itemize}
  \item no flattening - we use the relation given by \citet{Boogaard18} 
  in the entire mass range (extrapolating beyond the mass range
  7<log(M$_{\ast}$)<10.5 covered in their study when necessary)
  \item moderate flattening - we use the relation given by \citet{Boogaard18}
	at low masses and modify the slope at M$_{\ast}\geq 10^{9.7} \Msun$
	according to the results obtained by \citet{Speagle14} for  M$_{\ast}\geq 10^{9.7} \Msun$
	and 0<z<6. 
	In this variation the slope 
	of the high mass part of the SFMR is less steep than that of the low mass part and
	evolves with redshift, becoming steeper with increasing $z$.
	At z>6 we extrapolate the high mass-end slope evolution with time found by these authors.
  \item sharp flattening - we use the relation given by \citet{Boogaard18}
	at low masses and combine it with the relation obtained by \citet{Tomczak16}
	\footnote{
	\citet{Tomczak16} used a different parametrization of the SFMR (as opposed to the one given 
	by eq. \ref{eq: SFMR}):
	$\rm log\left( SFR \right) = s_{0} - log\left[ 1+ \left(\frac{ M_{\ast} }{ M_{TO;\ SFMR} }\right)^{-\gamma} \right]$
	,where $\gamma$ defines the low mass-end slope, $M_{TO;\ SFMR}$ is the stellar mass above which the
	relation begins to flatten and asymptotically approaches a peak value $s_{0}$
	}
	, which leads to flattening in SFMR at stellar masses above a 
	certain turnover mass M$_{\ast}\geq \rm M_{TO;\ SFMR}$.
	In this case SFR is almost constant with increasing M$_{\ast}$ at M$_{\ast}\geq \rm M_{TO;\ SFMR}$,
	as opposed to the `moderate flattening` variation, 
	in which the slope changes but the SFR is still increasing with increasing mass.
	\citet{Tomczak16} concluded that $M_{TO;\ SFMR}$ increases with redshift in the redshift range
	0.5<z<4 covered in their study. At z<0.5 (z>3.28
	\footnote{
	We use the value of $M_{TO;\ SFMR}$ at z=3.28 instead of that corresponding to the edge of 
	the last redshift bin (3<z<4) used in the study by \citet{Tomczak16}, because their polynomial fit 
	log($M_{TO;\ SFMR}$)=0.9458 + 0.865z - 0.132z$^2$ reaches maximum at z$\approx$3.28 
	and a further decrease in $M_{TO;\ SFMR}$ seems to be an artifact.
	}) 
	we fix $M_{TO;\ SFMR}$ to the value that results from
	the relation fitted by these authors at z=0.5 (z=3.28).   
 \end{itemize}
 
 \begin{figure}
\centering
\includegraphics[scale=0.46]{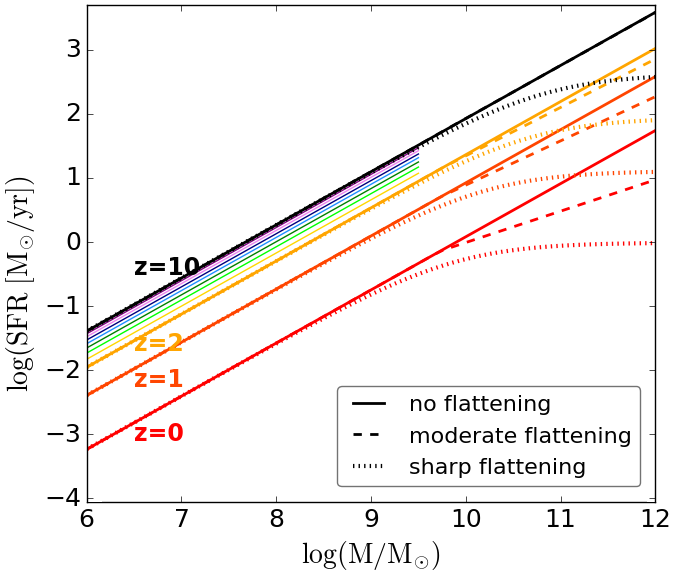}
\caption{ 
Star formation $-$ mass relation with three versions of the high-mass part shape:
no flattening (solid), moderate flattening \citep{Speagle14} and sharp flattening
\citep{Tomczak16} plotted at four example redshifts (z=0, z=1, z=2 and z=10, 
with the top lines corresponding to the highest redshift).
Thin solid lines plotted at the low mass part of the relation at z$\geq$3
indicate the evolution of the normalization of SFMR at 10$>$z$\geq$3, with 
spacing of $\Delta z$=1 in redshift between them.
}
\label{fig: SFRM_z}
\end{figure}
 In all variations we keep the normalization given by \citet{Boogaard18} 
 (converted to \citet{Kroupa01} IMF) at low masses and z$\sim$0
 and tie the high mass part of the relation to the low mass part accordingly.
 The normalization increases with redshift as $c\times \rm log(1+z)$
 with $c$=2.8 at z$\leq$1.8 and $c$=1 at z$>$1.8 (see Fig. \ref{fig: SFRM_z}).
 \\
 Similarly to the MZR, for any given stellar mass and redshift we 
 assume that the SFR is normally distributed around the mean given by the SFMR
 with $\sigma_{\rm SFR}$=0.3 dex.

\subsection{The model and its variations}

Table \ref{tab: model} summarizes the default choice of parameters used in our calculations.
The parameters that were varied are indicated in bold. We perform the calculations for 24 variations
of the parameters of the model. Each variation is defined by the choice of MZR, SFMR flattening and $\alpha_{fix}$.

 \begin{table*}
\centering
\small
\caption{
The default choice of parameters used in our calculations. The parameters that were varied 
are emphasized in bold.
}
\begin{tabular}{c c c }
\hline
 parameter & value/range &  description \\ \hline
  \hline
\multicolumn{3}{c}{ \textit{metallicity} } \\ \hline \hline
 $Z_{O/H}$ & 5.3 $-$ 9.7 &  12 + log(O/H) $-$ oxygen to hydrogen abundance ratio; binned, bin size = 0.022 \\ \hline
 $Z_{O/H \odot}$ ($Z_{\odot}$) & 8.83 (0.017) & solar metallicity \citep{AndersGrevesse89} \\ \hline
 \textbf{MZR} & T04/M09/KK04/PP04 & mass $-$ metallicity relation, varied: see Tab. \ref{tab: MZR} and Fig. \ref{fig: MZR_z} \\ \hline
 \multirow{3}{*}{$\sigma_{0}$} &    & scatter in the MZR (dispersion of the normal distribution):\\
		      & 0.1 dex & at $\rm log(M_{*}/\Msun)>9.5$\\ 
		      &$-$0.04 $\rm log(M_{*}/\Msun)$+0.48 dex  & at $\rm log(M_{*}/\Msun)\leq 9.5$\\ \hline
 $\sigma_{\nabla Z_{O/H}}$ & 0.14 dex (but see appendix \ref{app: gradients}) & dispersion of the normal distribution, $Z_{O/H}$ distribution within galaxies  \\ \hline
 \hline
\multicolumn{3}{c}{ \textit{star formation $-$ mass relation} } \\ \hline \hline
\textbf{SFMR flattening}&none/moderate/sharp & the high mass end slope of the star formation $-$ mass relation; varied: see Fig. \ref{fig: SFRM_z} \\ \hline
$a$ & 0.83 & slope of the SFMR at low masses \citep{Boogaard18} \\ \hline
\multirow{3}{*}{$c$} &  & redshift evolution coefficient of the SFMR (Sec. \ref{sec: method: SFMR-z})\\
		     & 2.8 & $z<1.8$ \\
		     & 1 & $z\geq$1.8 \\ \hline
$\sigma_{SFR}$ & 0.3 dex & scatter in the SFMR (dispersion of the normal distribution) \\ \hline
\hline
\multicolumn{3}{c}{ \textit{galaxy stellar masses} } \\ \hline \hline
log($M_{*}/\Msun$) & 6$-$12 & logarithm of stellar mass of galaxies \\ \hline
N$_{gal}$ & $10^{4}$ & number of bins in stellar mass (equally spaced in logarithm) \\ \hline
N$_{sampl}$ & 50 & number of galaxies sampled within each mass bin \\ \hline
$\pmb{\alpha_{fix}}$ &$-$1.45 or $-$0.1$\times \ z-$1.34 & low mass end slope of the GSMF; varied (see Fig. \ref{fig: GSMF_vs_z} and Fig. \ref{fig: alpha_GSMF_vs_z})\\ \hline
\hline
\multicolumn{3}{c}{ \textit{other} } \\ \hline \hline
IMF & K01; 0.1 $-$ 100 $\Msun$ & initial mass function \citep{Kroupa01} \\ \hline
$t$ & 464.4 $-$ 13465.7 Myr & time (age of the Universe), step size $\Delta t$= 100 Myr (if $\Delta z <$0.2) or calculated from $\Delta z$ \\ \hline
$z$ & 0 $-$ 10 & redshift, step size $\Delta z$=min( calculated from $\Delta$t or 0.2) \\ \hline
\end{tabular}
\label{tab: model}
\end{table*}

\section{Results}\label{sec: results}

One of the goals of this study is to find the observation-based 
distribution of the cosmic star formation rate density at different
metallicities and redshift SFRD(Z,z). 
Given the currently unresolved observational issues, 
especially the unknown source of the differences 
between the metallicities measured using different calibrations
and the shape of the high mass part of the SFMR (see sec \ref{sec: section2}), 
there is no simple, single answer to the question about the shape of the SFRD(Z,z).
Hence, instead of indicating one `best` SFRD(Z,z) we first choose an example `moderate`
variation of our model to discuss the general characteristics of the resulting distribution. 
We then discuss the differences between the variations allowed by the current observations and 
focus on the extreme cases that lead to the highest fraction of 
low/high metallicity star formation,
which delineate the uncertainty of our result.

\subsection{The distribution of the cosmic star formation rate at different
 metallicities and redshift: general picture}

 \label{sec: results: moderate}

\begin{figure*}
\centering
\includegraphics[scale=0.45]{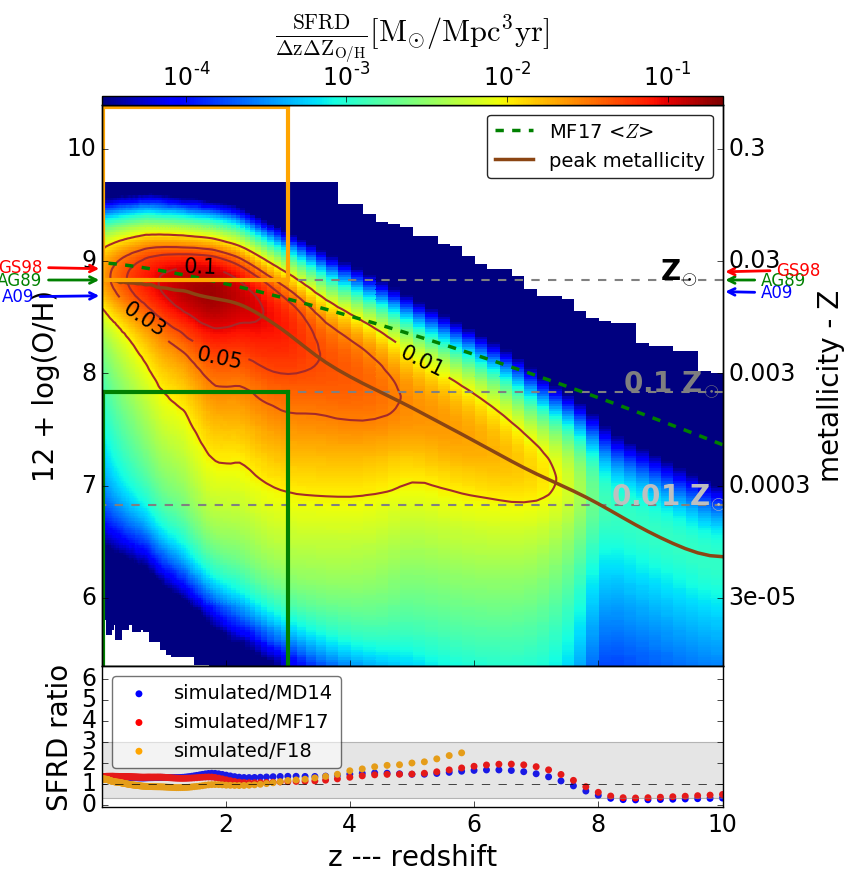}
\caption{
Upper panel: distribution of the star formation rate density (SFRD)
at different metallicities and redshift ($z$)
for the model described in Sec. \ref{sec: results: moderate}.
The right scale assumes solar metallicity from \citet{AndersGrevesse89} (green arrows).
The $Z_{O/H \odot}$ and $\Zsun$ estimates by \citet{Asplund09} (\citealt{GrevesseSauval98})
are shown with blue (red) arrows for the reference.
The colour indicates the amount of SFRD at each metallicity and $z$,
with most of the star formation happening
in the red region of the $z$ $-$ metallicity space.
The brown line indicates $Z$ corresponding to the peak of the distribution at each $z$ and
several contours of the constant SFRD.
The dashed green line shows the mass-averaged mean $Z$ estimate from \citet{MadauFragos17}
 (see Sec. \ref{sec: discussion: comparison}).
In Sec. \ref{sec: results: differences} we compare the model variations by comparing the fraction of stellar mass
formed above $Z_{O/H \odot}$ and below 10\% $Z_{O/H \odot}$ since $z$=3 
(i.e. within the orange and green rectangles).
Bottom panel: the ratio of the total SFRD($z$) (summed over metallicities) calculated in our model
and inferred from observations (blue $-$ MD14; \citet{MadauDickinson14},
red $-$ MF17; \citet{MadauFragos17},
orange $-$ F18; \citet{Fermi18}). A factor of three deviation from unity is indicated by the gray area.
}
\label{fig: SFRD_ZZ_z_103f_3}
\end{figure*}

Figure \ref{fig: SFRD_ZZ_z_103f_3} shows distribution of the star formation rate density 
at different metallicities and redshifts calculated assuming 
a moderate flattening at the high mass end of SFMR, 
M09 mass metallicity relation and fixed low mass end slope
of the galaxy stellar mass function $\alpha_{fix}$=$-$1.45.
We use this variation as an example to point out the main characteristics of the SFRD(Z,z)
that are common to all model variations:

\begin{itemize}
 \item At low $z$ the star formation is concentrated at relatively high metallicities;
 the location of the peak (thick brown line in Fig. \ref{fig: SFRD_ZZ_z_103f_3}) depends primarily on the MZR
\item The metallicity at which SFRD(z)\footnote{SFRD(Z,z) integrated over metallicities}
 peaks decreases towards higher $z$, reflecting the evolution of the MZR.
 The decrease rate increases around the redshift $\gtrsim$2 
 \footnote{At z$\gtrsim$2 the peak decreases almost linearly, 
 at a slightly higher rate than the normalization of the MZR
($-$0.28 dex versus d$Z_{O/H}$/dz = $-$0.26 dex in the case shown in Fig. \ref{fig: SFRD_ZZ_z_103f_3})}
   and depends on the MZR.
\item The global peak of the distribution is reached at $z\sim$1.8, 
      coinciding with the peak of the star formation history of the Universe,
      and at $Z$ that depends on the model variation (primarily MZR)
\item The distribution is asymmetric with respect to the peak metallicity,
      showing the extended tail at low $Z$ (see also Fig. \ref{fig: distr_103f_3}).
      The contribution from the tail to the total SFRD(z) at high $z$ 
      is noticeably higher in the variations with $\alpha_{fix}$=$\alpha_{fix}(z)$
      (see Fig. \ref{fig: SFRD_ZZ_z_113f_3}).
\end{itemize}
The bottom panel of Figure \ref{fig: SFRD_ZZ_z_103f_3} shows the redshift evolution of the
ratio of the total SFRD(z) in our simulations 
to the SFRD from observational studies by
\citet{MadauDickinson14}, \citet{MadauFragos17} an \citet{Fermi18}.
We note that the only way the total observed SFRD was used in our model is to
set the redshift at which we change the rate of redshift evolution of the SFMR normalization
\footnote{
The observational studies discussed in sec. \ref{sec: SFMR-z} 
show that this transition occurs somewhere between z$\sim$1$-$2
and here we choose $z$=1.8, which is an average redshift of the peak of the star formation history
of the Universe resulting from studies by \citet{MadauDickinson14}, \citet{MadauFragos17} an \citet{Fermi18}
}
.
Thus, this comparison is still a valid test for our model.
The total simulated SFRD shows a remarkable agreement with observations up to very high redshifts,
staying within a factor of 2$-$3 from the observed SFRD up to $z\sim$8 for all variations that assume
$\alpha_{fix}$=$-$1.45 (but not for the cases with $\alpha_{fix}$=$\alpha_{fix}(z)$, see Sec. \ref{sec: results: alpha_z}).
\\ \newline
Note that the model variations that differ in the assumptions about the
SFMR and/or GSMF lead to different total SFRD(z) (which sets the normalization of the SFRD(Z,z)).
In general, SFRD increases up to $z\sim 1.8$, decreases towards higher $z$ and declines sharply around $z\sim 8$.
This evolution is a result of the interplay between the evolving SFMR and GSMF,
mostly their normalizations.
The shape of both relations also changes with redshift, 
but it has a secondary effect on the evolution of the total SFRD 
(except for the cases with $\alpha_{fix}$=$\alpha_{fix}(z)$ at $z\gtrsim 4$).
At $z\sim$8 the normalization of the GSMF decreases abruptly (see fig. \ref{fig: GSMF_vs_z}),
which results in an apparent jump in the normalization in Figure \ref{fig: SFRD_ZZ_z_103f_3}.

\subsubsection{The variations with $\alpha_{fix}$=$\alpha_{fix}(z)$}\label{sec: results: alpha_z}

\begin{figure}
\centering
\includegraphics[scale=0.42]{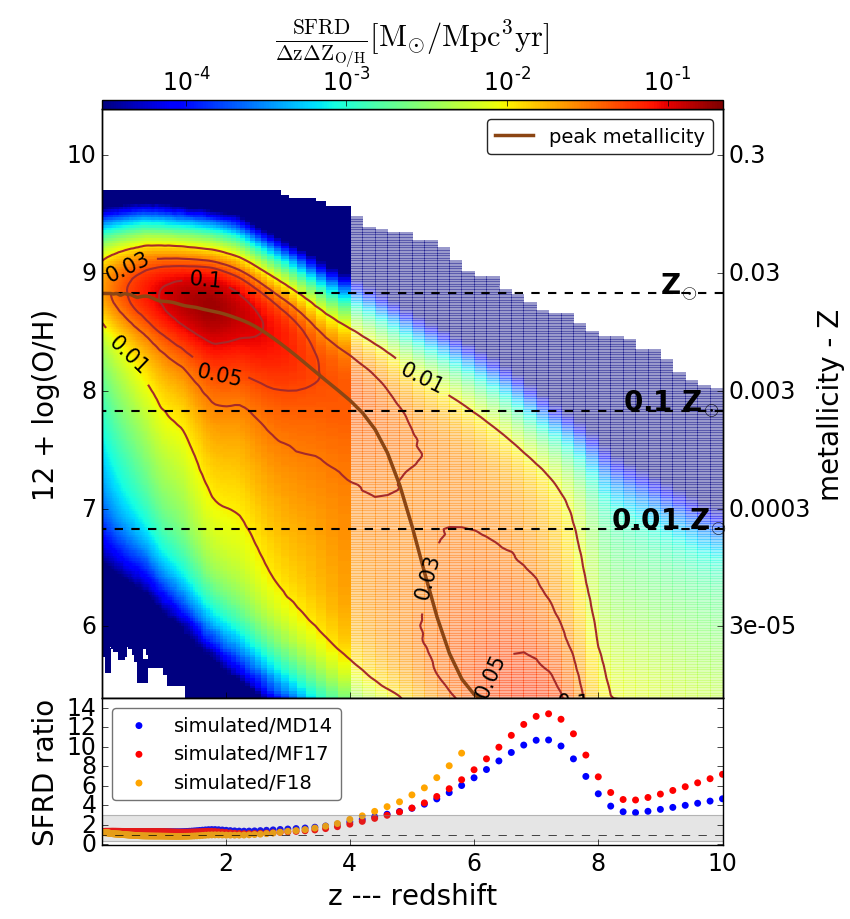}
\caption{ 
Same as Figure \ref{fig: SFRD_ZZ_z_103f_3}, but calculated
for the variation in which the low mass end slope of the GSMF 
evolves with redshift $\alpha_{fix}$=$\alpha_{fix}(z)$.
Note that at high redshifts z$\gtrsim$4 (shaded area) the
SFRD(z) is significantly higher than estimated from observations
(bottom panel; see Sec. \ref{sec: results: alpha_z} for the details).
This is true for all variations with $\alpha_{fix}$=$\alpha_{fix}(z)$.
Those variations were excluded from the analysis at z$\gtrsim$4.
For further description see the caption of Fig. \ref{fig: SFRD_ZZ_z_103f_3}.
}
\label{fig: SFRD_ZZ_z_113f_3}
\end{figure}

When the low mass end slope of the GSMF is allowed to evolve (steepen) with redshift according 
to the fit shown in Fig. \ref{fig: alpha_GSMF_vs_z}, the resulting SFRD(Z,z) distribution starts to
deviate significantly from the picture shown in Fig. \ref{fig: SFRD_ZZ_z_103f_3} at $z\gtrsim 4$.
The SFRD(Z,z) distribution calculated assuming the same relations as in the case
shown in Fig. \ref{fig: SFRD_ZZ_z_103f_3}, but with evolving $\alpha_{fix}$=$\alpha_{fix}(z)$
is shown in Figure \ref{fig: SFRD_ZZ_z_113f_3} (the other model variations with $\alpha_{fix}$=$\alpha_{fix}(z)$ 
show the same qualitative behavior).
\\
The normalization of the SFMR continuously increases with redshift, 
which means that galaxies at all masses produce stars at higher rates. 
At the same time, the normalization of the GSMF decreases and the cut off mass shifts
to smaller values, both reducing the number density of the most massive galaxies 
(with the highest SFR and $Z$ at a certain redshift). 
$\alpha_{fix}$ decreases (the slope steepens), increasing the number of low mass galaxies
and the shape of the GSMF evolves almost into a single power-law (see fig. \ref{fig: GSMF_vs_z}).
This, together with continuously increasing SFR(M$_{\ast}$) produces a plateau in the total SFRD(z)
in our simulations at z$\sim$4-7 (see e.g. fig. \ref{fig: CCSN rate comparison},
 showing the core collapse supernovae rate, which is proportional to the total SFRD(z)). 
The low mass galaxies begin to dominate the total star formation budget and the peak of
the SFRD(Z,z) shifts to very low metallicities (below 1\% $\Zsun$ at $z\sim$5).
\\
Such plateau in SFRD(z) at z$\sim$4$-$7 is not found within the current observational studies 
(predicting a continuous decrease in the total SFRD(z) at these high redshifts)
and hence the bottom panel of Fig. \ref{fig: SFRD_ZZ_z_113f_3} shows that our simulations
start to significantly overpredict the total SFRD at z$\sim$6. 
\\
Taking this into account, we consider the high redshift ($z\gtrsim 4$) 
predictions of the model variations with $\alpha_{fix}$=$\alpha_{fix}(z)$ unreliable
and exclude them from further analysis, while still taking them into consideration at lower redshifts.
We return to that discrepancy at high redshifts in Sec. \ref{sec: discussion: high-z}.

\subsection{The extreme cases} \label{sec: results: comparison}

The main purpose of this study is to identify the model variations 
 that lead to the most \emph{extreme}
pictures of the SFRD(Z,z) and hence delineate its uncertainty.\\
Different assumptions about the amount of star formation occurring
at low metallicities can lead to significant differences e.g. in the obtained properties of
the progenitors of merging double black holes and long gamma ray bursts and the inferred rates
of transients connected to those objects.
Keeping in mind the importance of metallicity,
we interpret extremes as variations that lead to the 
smallest and the highest fraction of stellar mass forming at low/high metallicity.\\
The main differences in the SFRD(Z,z) distributions obtained for different
model variations are the following:
\begin{itemize}
 \item The location of the peak metallicity and the slope of the curve indicating
       the peak metallicity at each redshift
 \item The extent of the low metallicity tail and its contribution to the total SFRD(z)
 \item Normalization of the distribution / total SFRD(z) 
\end{itemize}
Focusing on the fraction of stellar mass formed at \emph{low} and \emph{high metallicity}
means that we set aside the differences in normalizations (but see Sec. \ref{sec: results: total mass}).
\\ 
The location of the peak metallicity and the rate of its decrease with $z$ are set primarily by the choice of the MZR
-- the relation with high, slowly decreasing normalization will maximize the fraction of stellar mass forming at high metallicity. 
The slope of the MZR affects the low metallicity tail of the SFRD(Z,z) - the flatter the relation,
the smaller the difference in metallicity of stars forming in galaxies of different masses
and the the low-Z tail is reduced.
\\
Furthermore, the choice of the SFMR regulates the contribution of the most massive galaxies (forming stars at high metallicities)
to the total SFRD(z). Hence, the SFMR with no flattening maximizes the high metallicity star formation,
and the reverse is true for the SFMR with sharp flattening
\\
Finally, the low mass end slope of the GSMF regulates the contribution of galaxies with the lowest masses.
The steeper the slope, the more low mass galaxies 
(and hence more stars forming in the low metallicity tail of the distribution).
\\
Thus, one can expect the extreme variations to be represented by the combination of
$KK04$ MZR, SFMR with no flattening and $\alpha_{fix}$=$-$1.45 (\emph{the high metallicity extreme})
and $PP04$ MZR, SFMR with sharp flattening
\footnote{
The choice of the $\alpha_{fix}$ variation in the low metallicity extreme is not obvious.
However, as can be seen by comparing Fig. \ref{fig: SFRD_ZZ_z_103f_3} and \ref{fig: SFRD_ZZ_z_113f_3},
the choice of the $\alpha_{fix}$ variation affects the SFRD(Z,z) mostly at $z\gtrsim$3, up to which
redshift both variations give similar results.
At the same time, the variations with $\alpha_{fix}$(z) are excluded at $z\gtrsim$4 (Sec. \ref{sec: results: alpha_z})
and we focus on the discussion of the variations with $\alpha_{fix}$=$-$1.45. See Sec. \ref{sec: results: differences}.
}
(\emph{the low metallicity extreme}).

\subsubsection{SFRD(Z,z) for the extreme cases}\label{sec: results: extreme}

\begin{figure*}
\centering
\includegraphics[scale=0.43]{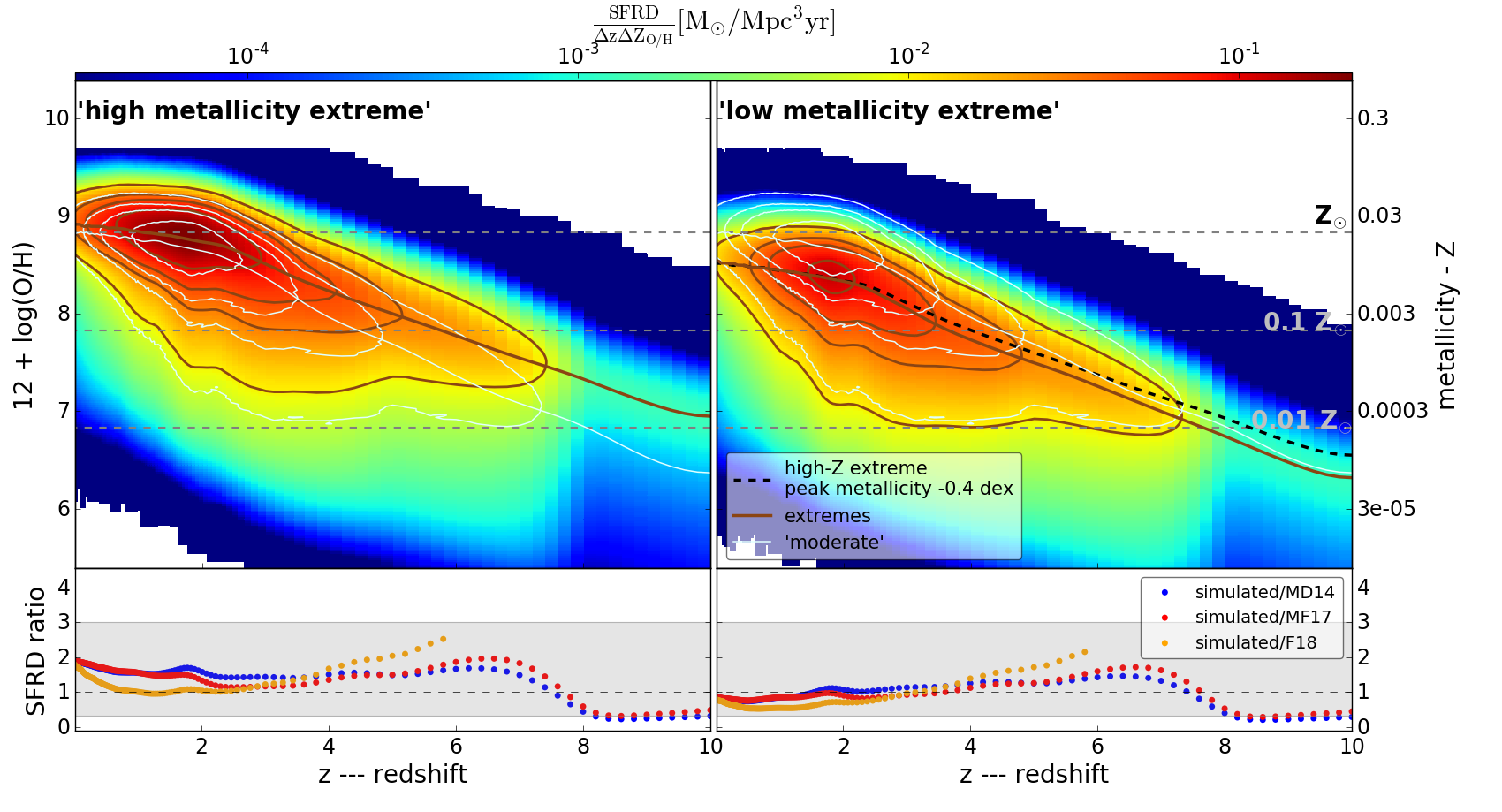}
\caption{
Upper panels: distribution of the star formation rate density (SFRD) at different metallicities and redshift ($z$).
Left -- the model variation that leads to the highest fraction of mass formed at high metallicities,
right -- the model variation that leads to the highest fraction of mass formed at low metallicities.
Difference between the two extremes depicts the uncertainty of the SFRD(Z,z) based on the observational 
properties of star forming galaxies. \newline
The thick brown line indicates the metallicity corresponding to the peak of the SFRD(z)
and the contours of constant SFRD (with the same levels as in Fig. \ref{fig: SFRD_ZZ_z_103f_3}).
We overplot the contours calculated for the moderate variation discussed in Sec. \ref{sec: results: moderate}
and plotted in Fig. \ref{fig: SFRD_ZZ_z_103f_3} (thin white line).
The thick dashed black line in the right panel shows the peak metallicity from the left panel,
downshifted by 0.4 dex to match the metallicities of both extremes at $z\sim1.8$.
Bottom panels: the ratio of the total SFRD(z) calculated in the two variations 
of our model and as inferred from observations. 
For further description see the caption of Fig. \ref{fig: SFRD_ZZ_z_103f_3}
}
\label{fig: SFRD_ZZ_z_extremes}
\end{figure*}

The distributions of the star formation rate density across metallicities
and redshifts for the extreme cases identified in Sec. \ref{sec: results: comparison}
are shown in Figure \ref{fig: SFRD_ZZ_z_extremes}. 
The difference between the two panels visually demonstrates the uncertainty of the SFRD(Z,z)
due to unresolved questions in the determination of various characteristics of galaxies 
(absolute metallicity scale, high mass end of the SFMR and low mass end of the GSMF).
\\ \newline
To facilitate the comparison, we overplot the constant SFRD contours and the line
showing peak metallicity at each redshift for the moderate variation shown in Fig. \ref{fig: SFRD_ZZ_z_103f_3}.
The left panel of Figure \ref{fig: SFRD_ZZ_z_extremes} shows the high metallicity extreme.
The low metallicity extreme is shown in the right panel.
\\
As shown in the right panel in Fig. \ref{fig: SFRD_ZZ_z_extremes}, 
except for the $\sim$0.4 dex offset, the redshift evolution of 
the peak metallicity for both extremes is very similar.
This stems from the fact that MZRs obtained for different metallicity calibrations
have similar shape (i.e. are relatively flat) in the high mass part 
that contributes the most to the total star formation.
At higher redshifts $z\gtrsim 2$ the peak metallicity lines for both cases start to deviate
due to different rates at which the MZR normalization decreases for both relations
(the PP04 MZR normalization decreases at a higher rate than KK04; see Table \ref{tab: MZR}).
The peak metallicity reaches our conventional low metallicity limit of 0.1$\Zsun$
at z=5.8 and z=3.6 for the high and low metallicity extreme respectively
and never drops below  0.01$\Zsun$ in the former case.
\\
Note that the low metallicity extreme 
has a lower SFRD(z) and leads to smaller total stellar mass than the moderate variation
and the high metallicity extreme (see Sec. \ref{sec: results: total mass}).

\subsubsection{Quantifying the differences between the variations}\label{sec: results: differences}
In this section we aim to quantify the differences between the model variations discussed above,
in particular to find the maximum difference between the fraction of stellar
mass formed at \emph{low} and \emph{high metallicity} across the considered variations.
\\
For presentation purposes we adopt the conventional limits of 
$Z<$0.1$\Zsun$ and $Z>\Zsun$ to define the low and high metallicity respectively
 \citep[assuming solar abundances according to][]{AndersGrevesse89}.
 We note that this particular choice does not affect our conclusions.
 \\
 In our analysis we focus on three redshifts ranges: $z<0.5$ to which we refer as the `local Universe`
and which can be probed with the current network of ground-based gravitational wave detectors, 
$z<3$ up to which redshift our model is still reasonably well backed up by the current observational
results and which captures the great majority of the star formation history of the Universe
and $z<10$ where we extend our calculations beyond redshifts directly probed in observational
studies relying on extrapolations.
Those redshifts correspond to $\sim$12.8 Gyr,  $\sim$2.1 Gyr and $\sim$0.46 Gyr 
respectively in terms of the age of the Universe.
The example distribution of stellar mass formed since $z$=3 across metallicities
is shown in appendix \ref{app: peak}.
\\ \newline
Figure \ref{fig: lowZ vs highZ SFRD} shows the fraction of stellar mass formed at low metallicities 
versus that formed at high metallicities since $z$=0.5, 3 and 10 for our model variations.
It can be seen that, as expected, variations leading to the highest fraction of mass formed at low metallicities
at the same time produce the lowest fraction of high metallicity stars (and vice versa) at all redshifts
and the extreme cases introduced earlier in Sec. \ref{sec: results: comparison} can be easily identified
(low-Z extreme $-$ brown circles; high-Z extreme $-$ red squares).
Note that when the stellar mass formed since $z>1.5$ is considered, 
the fraction of stellar mass formed at low-Z is additionally increased 
if the low mass end slope of the GSMF is allowed to steepen with $z$
\footnote{
At redshifts $z<1.1$ $\alpha_{fix}(z)$ resulting from the fit shown in Fig. \ref{fig: alpha_GSMF_vs_z}
is bigger than $\alpha_{fix}$=$-$1.45 that we use in the variations in which the slope does not evolve with $z$.
Hence, there are less low mass galaxies in the variations with $\alpha_{fix}(z)$ in that redshift range and
a smaller fraction of stellar mass forms at low metallicities. For the variation with 
PP04 MZR and sharp flattening in the SFMR those fractions equalize at $z\sim1.5$.
} (compare brown circles with and without edge).
The impact of different assumptions on the low-$Z$ mass fraction is further discussed in appendix \ref{app: lowZ mass fraction}.
\\ \newline
The difference between the fraction of the mass formed at low and high metallicity
produced in the two extremes is quite significant.
Those values (summarized in Table \ref{tab: Mtot})
range between $\sim$3\%$-$15\% and $\sim$4\%$-$55\% at $z\leq$0.5,
$\sim$9\%$-$27\% and  $\sim$1\%$-$27\% at $z\leq$3 and
$\sim$17\%$-$43\% and $\sim$0.7\%$-$19\% at $z\leq$10 for the low and high metallicity fractions respectively.
\\
For the example variation shown in Fig. \ref{fig: SFRD_ZZ_z_103f_3}
(with M09 MZR, moderate SFMR flattening and non-evolving $\alpha_{fix}$)
the corresponding values are 6\% and 38\% at $z\leq$0.5,
15\% and 15\% at $z\leq$3, 
and  29\% and 11\% at $z\leq$10.
We note that it does not fall exactly in between the low and
high metallicity star formation extremes,
but produces a higher fraction of mass at high metallicities with respect to a simple average.
While it represents the `most moderate` case considered in this study 
(and hence we refer to it as \emph{the moderate variation}), 
the results obtained for the T04 MZR (for the same assumptions about the SFMR and GSMF) are comparable.
This highlights the question of degeneracies involved in our calculations.
Similar results could be obtained using another MZR (resulting from a different metallicity calibration),
or by combining it with different assumptions about the SFMR and/or GSMF 
that fall between the extreme cases considered in our work.
\\ \newline
In nearly all cases, the summed fraction of stellar mass produced at low and high metallicity is smaller than
0.5. This means that in those variations most of the stellar mass forms outside the assumed low and 
high metallicity regimes
(i.e. between 0.1$\Zsun$ and $\Zsun$ or, equivalently, between $Z_{O/H}$=7.83 and $Z_{O/H}$=8.83)
\footnote{
If only the local redshift range is considered, the exceptions are the variations with KK04 and
no or moderate SFMR flattening and T04 MZR with no SFMR flattening. 
We find that at $z<0.5$ in those cases more than half of the stellar mass
formed at metallicities higher than the solar value}.
We look in more detail at the location of the peak of the distribution
- i.e. metallicity at which most of the stellar mass formed in different redshift ranges
and model variations in appendix \ref{app: peak}.

\begin{figure*}
\centering
\includegraphics[scale=0.4]{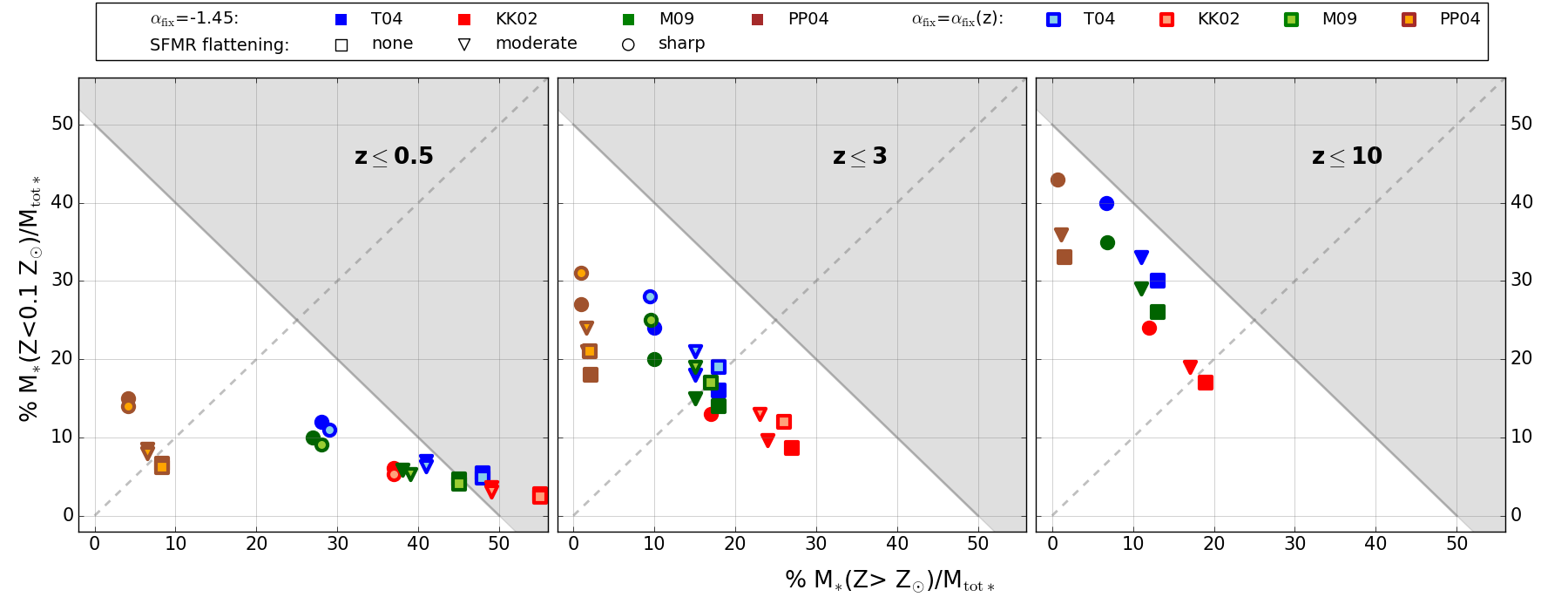}
\caption{ 
The fraction of mass formed in stars at low ($<0.1\Zsun$) vs at high ($>\Zsun$) metallicities 
since redshift z=0.5 (left), z=3 (middle) and z=10 (right) for different model variations.
Different symbols distinguish the assumptions about the high mass part of the SFMR
while the colours distinguish the assumptions about the MZR and low mass end slope of the GSMF.
We exclude the variations with $\alpha_{fix}(z)$ at z$\gtrsim$4 (see Sec. \ref{sec: results: alpha_z}).
The results shown in Fig. \ref{fig: SFRD_ZZ_z_103f_3} (\ref{fig: SFRD_ZZ_z_113f_3}) 
correspond to the variation indicated with the dark (light) green triangle. 
The extreme cases discussed in Sec. \ref{sec: results: extreme} are
indicated with the dark red square and brown/orange circle.
The dashed diagonal line corresponds to equal percent of $M_{*}$ formed at low and high $Z$.
Almost all points fall outside the gray shaded area, which means that more than 50\% of $M_{*}$ 
formed at metallicities between 0.1 $\Zsun$ and $\Zsun$.
}
\label{fig: lowZ vs highZ SFRD}
\end{figure*}

\subsection{Total stellar mass vs the mass fraction}\label{sec: results: total mass}

 \begin{table}
\centering
\small
\caption{
Total stellar mass (column 3) formed since redshift $z_{i}$ (column 1)
in different versions of the model and fraction
of that mass formed at low (column 4) 
and high (column 5) metallicity.
The second column specifies the shape of the high mass end of the SFMR.
The lowest (highest) value given in the column 4 (5)
corresponds to the KK04 MZR.
The highest (lowest) value given in the column 4 (5)
corresponds to the PP04 MZR.
$z_{i}$=10 was not included for the $\alpha_{fix}$=$\alpha_{fix}(z)$ 
due to the reasons discussed in Sec. \ref{sec: results: alpha_z}.
}
\begin{tabular}{c c c c c}
\hline
 z$_{i}$ & SFMR flattening & $\frac{M_{tot \ast}(z\leq z_{i})}{\Msun}$ 
 & $\frac{\rm M_{\ast}(Z<0.1 \Zsun)}{\rm M_{tot \ast}}$ &$\frac{\rm M_{\ast}(Z> \Zsun)}{\rm M_{tot \ast}}$\\ \hline \hline
\multicolumn{5}{c}{ $\alpha_{fix}$=$-$1.45 } \\ \hline \hline
 0.5  &        & 5.1$\times 10^{18}$ & 0.03-0.07 & 0.08-0.55\\
 3    & none   & 5.4$\times 10^{20}$ & 0.09-0.18 & 0.02-0.27\\
 10   &        & 7.6$\times 10^{20}$ & 0.17-0.33 & 0.01-0.19\\ \hline
 0.5  &        & 2.3$\times 10^{18}$ & 0.06-0.15 & 0.04-0.37\\
 3    & sharp  & 3.6$\times 10^{20}$ & 0.13-0.27 & 0.01-0.17\\
 10   &        & 5.4$\times 10^{20}$ & 0.24-0.43 & 0.01-0.12\\ \hline
 0.5  &        & 4$\times 10^{18}$   & 0.04-0.09 & 0.07-0.49\\
 3    &moderate& 4.8$\times 10^{20}$ & 0.10-0.21 & 0.02-0.24\\
 10   &        & 6.9$\times 10^{20}$ & 0.19-0.36 & 0.01-0.17
 \\ \hline \hline
 \multicolumn{5}{c}{ $\alpha_{fix}$=$\alpha_{fix}(z)$ } \\ \hline \hline
 0.5  &        & 5.1$\times 10^{18}$ & 0.02-0.06 & 0.08-0.55\\
 3    & none   & 5.6$\times 10^{20}$ & 0.12-0.21 & 0.02-0.26 \\ \hline
 0.5  &        & 2.3$\times 10^{18}$ & 0.05-0.14 & 0.04-0.37\\
 3    & sharp  & 3.8$\times 10^{20}$ & 0.17-0.31 & 0.01-0.17\\ \hline
 0.5  &        & 3.9$\times 10^{18}$ & 0.03-0.08 & 0.06-0.49 \\
 3    &moderate& 4.9$\times 10^{20}$ & 0.13-0.24 & 0.02-0.23
 \\ \hline
\end{tabular}
\label{tab: Mtot}
\end{table}

The different variations do not only lead to different fractions of stellar mass formed at high and low metallicity, 
but also produce different total stellar mass (see Table \ref{tab: Mtot}).
Variations with no flattening in the SFMR lead the highest total $M_{*}$ because they
produce more stars in the high-mass galaxies (with high SFR and $Z$).
\\
This effect reduces the difference in the total mass formed at low metallicity 
and increases the difference in terms of 
the total mass formed at high metallicity between the variations with different assumptions about the SFMR.
\\
Using the values given in Table \ref{tab: Mtot}, we find that
the total mass formed at low metallicity across the model variations differs by less than a factor of 3
and is relatively well constrained within our study.\\
The corresponding values of the mass formed at high metallicity 
span a much wider range and the differences between the model variations reach up to a factor of $\sim$40.

\section{Application of the model: core collapse supernovae rates}\label{sec: CCSN rates}

\begin{figure*}
\centering
\includegraphics[scale=0.45]{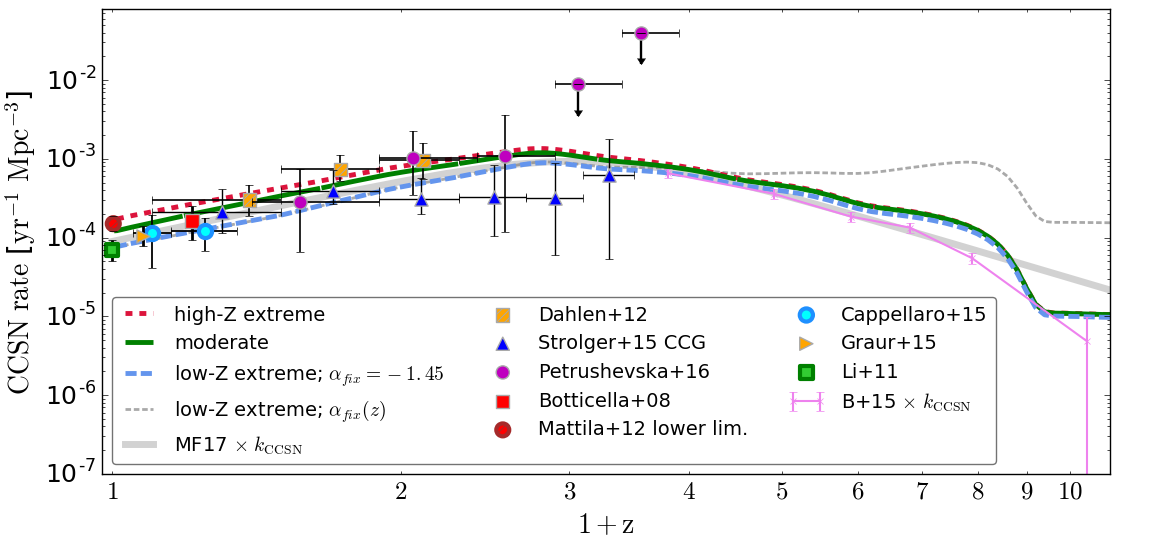}
\caption{ 
Core-collapse supernovae (CCSN) rate calculated using our model 
(short dashed red line $-$ high metallicity extreme; solid green line $-$ moderate; 
thick blue long dashed line $-$ low metallicity extreme; thin gray dashed line $-$ low metallicity extreme with
$\alpha_{fix}$ evolving with redshift) compared with the observational estimates
\citep[][see legend]{Botticella08,Li11_III,Dahlen12,Mattila12,Cappellaro15,Graur15,Strolger15,Petrushevska16}.
In case of \citet{Strolger15} we plot the CANDLES+CLASH+GOODS based estimate (Strolger+15 CCG)
as given by the authors and include the `extinction limits` systematic uncertainty (see Tab. 3 and 4 therein).
In case of the model results we assume that CCSN progenitors come from the initial mass range 8-25 $\Msun$.
For the reference, we also plot the SFRD(z) from \citet{MadauFragos17} (MF17, thick gray solid line)
and the high redshift SFRD estimate by \citet{Bouwens15} (B+15, magenta line; 
converted to our default \citet{Kroupa01} IMF) 
multiplied by the efficiency of formation of CCSN progenitors 
($k_{CCSN} \sim 0.009 \ \Msun^{-1}$; assuming CCSN mass range 8$-$25 $\Msun$ and Kroupa 2001 IMF).
}
\label{fig: CCSN rate comparison}
\end{figure*}

\begin{figure}
\centering
\includegraphics[scale=0.42]{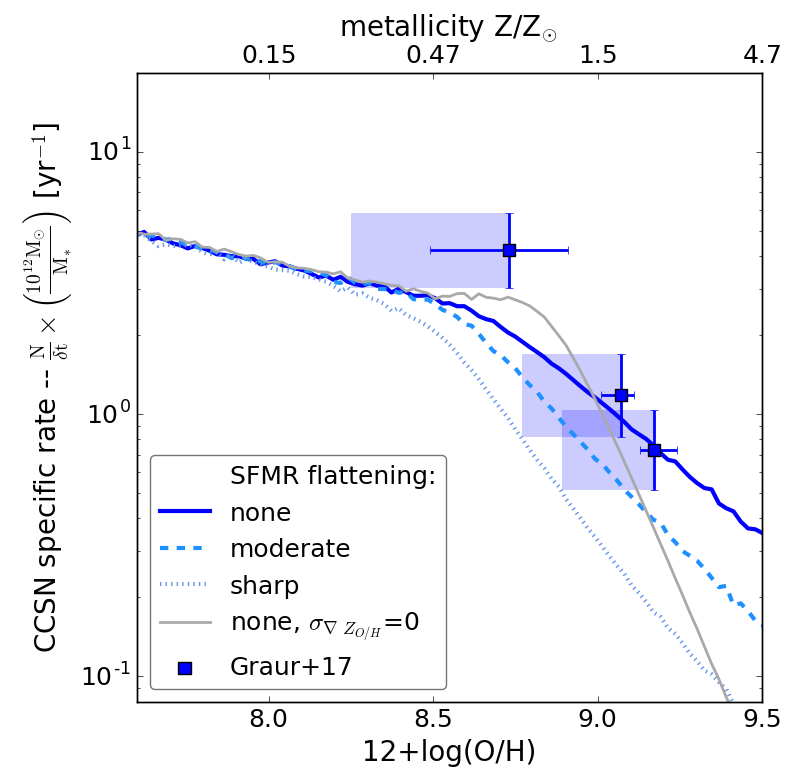}
\caption{ 
The specific CCSN rate as a function of metallicity 
for different model variations 
(SFMR with no flattening -- thick solid line, moderate flattening -- dashed line, sharp flattening -- dotted line)
and observations (\citet{Graur17}$^{18}$).
The metallicity scale assumes the T04 calibration$^{19}$.
The thin solid line shows the case with no flattening and without taking into account
the metallicity distribution within galaxies.
The blue shading shows the additional uncertainty ($-$0.24 dex) due to the fact that
the $Z_{O/H}$ in the observed sample was measured in the central region of the SN host
galaxy and not at the SN location.
Our model shows satisfying agreement with the data, although the
variations with sharp flattening in SFMR may underestimate the local high-Z star formation.
}
\label{fig: specific rate metallicity}
\end{figure}

Core-collapse supernovae (CCSN; i.e. Type II and Ib/c SNe) originate from short-lived massive stars
and can serve as an independent tracer of the star formation \citep[e.g.][]{MadauDickinson14}.
We can thus confront the predictions of our model with the volumetric CCSN rate estimates at different redshifts,
assuming that the volumetric CCSN rate is proportional to the SFRD(z).
However, one should keep in mind that there is a number of factors that make this comparison not straightforward
and may cause deviations from the simple proportionality (see appendix \ref{app: CCSN rates}).
\\
In Figure \ref{fig: CCSN rate comparison} we plot the observational CCSN rate estimates at different redshifts
against the star formation rate density for the two extreme cases and the moderate variation of our model
discussed in Sec. \ref{sec: results}, multiplied by a constant factor $k_{CCSN}\sim 0.009 \ \Msun^{-1}$.
This factor comes from a simple IMF-based scaling and corresponds to the 
efficiency of formation of CCSN progenitors assuming our default \citet{Kroupa01} IMF and CCSN forming
form single stars with masses between 8$-$25 $\Msun$ 
\footnote{
The $k_{CCSN}$ for z$\lesssim$1 and the moderate variation of our model
lies between $k_{CCSN}\sim$0.0082 $-$ 0.0097 $\Msun^{-1}$
(depending on the exact redshift range and data sample considered in the fitting).
Assuming a fixed low mass end of 8 $\Msun$ and our default \citet{Kroupa01} IMF, 
this corresponds to the upper CCSN progenitor mass end between $\sim$20 $-$ 30 $\Msun$.
}
, which given the large scatter in the observations,
provides a satisfying match to the data.
\\ \newline
The knowledge of the distribution of the cosmic star formation over metallicities
allows us to compare not only with the global volumetric CCSN rate discussed above,
but also to perform a more detailed comparison with the CCSN rates measured as a function of metallicity.
We calculate the specific CCSN rate as a function of metallicity
(number of CCSN in a certain metallicity bin divided by the stellar mass
in galaxies in that metallicity bin)
and use the mass range for the CCSN progenitors estimated based on the comparison with the volumetric rates
to get the normalization.
Those rates can be confronted with the specific CCSN rates based on the Lick Observatory Supernova Search 
(LOSS) sample from \citet{Graur17}
\footnote{
The CCSN specific rates were obtained by summing the specific rates for type II and stripped envelope SN.
The rates given in \citet{Graur17} were estimated using different metallicity bins for the two types of SN. 
Here we use three metallicity bins (in the Tremonti et al. (2004) scale): 
$8.71^{+0.18}_{-0.24}$, $9.07^{+0.04}_{-0.06}$, $9.17^{+0.07}_{-0.04}$ 
in which the corresponding 
type II SN rates are: $3.47^{+1.06}_{-0.83}$, $0.90^{+0.28}_{-0.22}$, $0.52^{+0.15}_{-0.12}$
and stripped envelope SN rates are: $0.97^{+0.66}_{-0.42}$, $0.34^{+0.27}_{-0.16}$, $0.25^{+0.17}_{-0.11}$
(Graur, private communication)
}.
Figure \ref{fig: specific rate metallicity} shows the specific rates at z$\lesssim$0.02, which roughly corresponds
to the redshift range relevant for the LOSS SN sample. We show the results for the model variations 
using the T04 MZR, as this was the metallicity calibration used in \citet{Graur17}.
\footnote{
We note that the choice of the MZR does not affect the result of the comparison.
To compare the observations with the results obtained for the model variation with a different MZR,
one has to convert the data points to a different metallicity calibration.
Applying the conversion method described in \citet{KewleyEllison08} to the observations
results in a shift in the metallicity of the data points which is consistent with the shift of
the model curves obtained for the different MZR and hence the conclusions drawn from the comparison do not change.
}
\\
The metallicity measurements used in \citet{Graur17} are based on the SDSS observations and 
correspond to the central metallicity within the CCSN host galaxy.
The central metallicity is typically higher than the metallicity at the CCSN location 
(due to radial metallicity gradients, see Sec. \ref{sec: MZR: gradients}).
It is also higher than the metallicity probed by the MZR, which gives the `averaged` global value
(and corresponds roughly to the metallicity that would be measured at the distance of one effective radius r$_{e}$).
Therefore, an offset in metallicity between the data and the results from our model can be expected.
To be conservative in our comparison, we extend the horizontal errorbars of the data 
shown in Figure \ref{fig: specific rate metallicity} by 0.24 dex towards lower metallicity,
assuming that the great majority of the observed CCSN explosion site metallicities would 
be contained within that range (see appendix \ref{app: SN location Z}).
This wide range, together with the uncertain translation of the
amount of star formation to the number of core collapse supernovae (which affects the location
of the model curves on the vertical axis) precludes us from drawing any strong conclusions from that comparison.
In general, we find a reasonably good agreement of our model with the data.
However, we note that the dotted curve in Fig. \ref{fig: specific rate metallicity} 
which corresponds to the model variation with the sharp flattening in the high mass end of the SFMR, 
falls slightly outside the extended uncertainty regions of the data points.
Hence, the model variations with a sharp flattening in the SFMR
may underestimate the amount high metallicity star formation in the local Universe.
\\
\newline
We note that the slope of the high metallicity part of the relation plotted in 
Fig. \ref{fig: specific rate metallicity}
depends on the combination of the scatter present in the MZR ($\sigma_{0}$) and on the 
width of the distribution of metallicities at which the stars form within galaxies 
($\sigma_{\nabla O/H}$; see appendix \ref{app: specific CCSN rate slope}).
The slope cannot be directly compared with the data from \citet{Graur17}, 
as they only provide the central metallicities of CCSN host galaxies,
but could potentially be probed with similar observations that provide metallicity measurements at SN location.

\section{Discussion}\label{sec: discussion}
In this section we further comment on the reliability of our results at high redshifts 
and on the adopted extrapolations.
We also discuss some of the simplifications present in our method.
Finally, we place our results in the context of previous studies and
perform a brief comparison with the predictions of cosmological simulations.

\subsection{Reliability of our results at high redshifts}\label{sec: discussion: high-z}

The current observational estimates of the GSMF used in our study reach to
$z\sim 9$ and the rate of redshift evolution of the SFMR
was also constrained at these high redshifts \citep[e.g.][]{Bhatawdekar18}.
However, the shape and evolution of the gas-phase MZR is essentially
unconstrained above z$\sim$3.5 with the current observations (see Sec. \ref{sec: MZR: evolution})
and our results at higher redshifts need to be taken with a grain of salt.
\\ 
The observational estimates of 
the total cosmic SFRD \citep[reaching even up to $z\sim10$ e.g.][]{Bouwens15}
can serve as constraints to our model,
allowing to validate the combined assumptions about the SFMR and GSMF.
SFRD(z) recovered from our model shows a reasonable agreement
with the observations up to z$\gtrsim$8 for all variations in which
the low mass end slope of the GSMF for star forming galaxies does not evolve with redshift
($\alpha_{fix}=-1.45$).
\\
However, there seems to be overall trend for $\alpha_{fix}$ to steepen with redshift 
(see Sec. \ref{sec: intro:GSMF} and Fig. \ref{fig: GSMF_vs_z}).
When we allow $\alpha_{fix}$ to evolve according to the fit presented in Fig. \ref{fig: GSMF_vs_z},
the cosmic SFRD(z) starts to significantly deviate from the observational estimates around $z\gtrsim$4
(see Sec. \ref{sec: results: alpha_z}).
The cause of this discrepancy is not clear.
\\
One of the assumptions which may be responsible for this problem
is that the low-mass end slope of the SFMR ($a$=0.83) does not evolve with redshift. 
As discussed in Sec. \ref{sec: SFMR-z}, it is difficult to constrain the evolution of $a$
with the current observations.
If the value of $a$ would increase with redshift,
the amount of star formation occurring in low mass galaxies could be reduced and
this would alleviate the tension with the cosmic SFRD. 
Such evolution has been reported by several authors \citep[e.g.][]{Whitaker14,Pearson18},
while some other studies observe the reverse trend \citep[e.g.][]{Kurczynski16,Bisigello18}.
Another argument in favor of the steepening $a$ comes from the studies 
comparing the evolution of the stellar mass buildup resulting from 
the SFMR and as is observed from the evolution of the GSMF \citep[e.g.][]{Weinmann11,Leja15}.
The observationally inferred SFMR is known to lead to an overabundance of the low mass galaxies
with respect to the GSMF even when the high merger rate 
(shifting some of the excess galaxies to higher masses) is assumed for those galaxies.
\citet{Leja15} show that $a\sim$1 at low masses 
(especially at the high redshift end of the redshift range considered in their study z$\lesssim$2 ) 
helps to improve the consistency (although does not entirely solve the problem).
\\
The inconsistency found in our variations with $\alpha_{fix}$=$\alpha_{fix}(z)$
may also arise from the assumed extrapolations to low stellar masses.
We extend the scaling relations and GSMF down to $M_{\ast}=10^{6}\Msun$, which is much lower
than the stellar mass that can be probed by any high-redshift survey.
In principle, there is no guarantee that the slope of the GSMF and SFMR does not change
at stellar masses below the range probed by the current observations or that such a turnover
appears at high redshifts.
In fact, recent studies by \citet{Bouwens17} and \citet{Atek18} investigating 
the galaxy luminosity function (LF) in the extremely faint regime
with lensed galaxies find support for the presence of a flattening/turnover in the LF
at $z\sim$6 and absolute UV luminosities $M_{UV}\gtrsim -15$ mag
\footnote{
According to the mass - luminosity relation at z=6 from \citet{Song16},
$M_{UV}\sim -15$ mag and $M_{UV}\sim -14$ mag would correspond to
log$M_{\ast}\sim$6.5 and log$M_{\ast}\sim$6 respectively
and hence is close to the lower limit of the stellar mass of galaxies
considered in our study.
}.
The upper limit on the UV emissivity at z$\sim$5$-$6 found by \citet{Fermi18}
also favors the presence of a turnover in the LF at $M_{UV}\gtrsim -14$ mag.
However, the flattening/turnover at these low luminosities is not likely to solve
the discrepancy observed in our model.
\\
Finally, we note that systematic trends in the IMF (e.g. with metallicity/redshift), if present,
could have a strong and non-straightforward effect on our results (see Sec. \ref{sec: IMF}).

\subsection{Comparison with the Illustris cosmological simulations}\label{Sec: discussion: Illustris1}
\begin{figure*}
\centering
\includegraphics[scale=0.33]{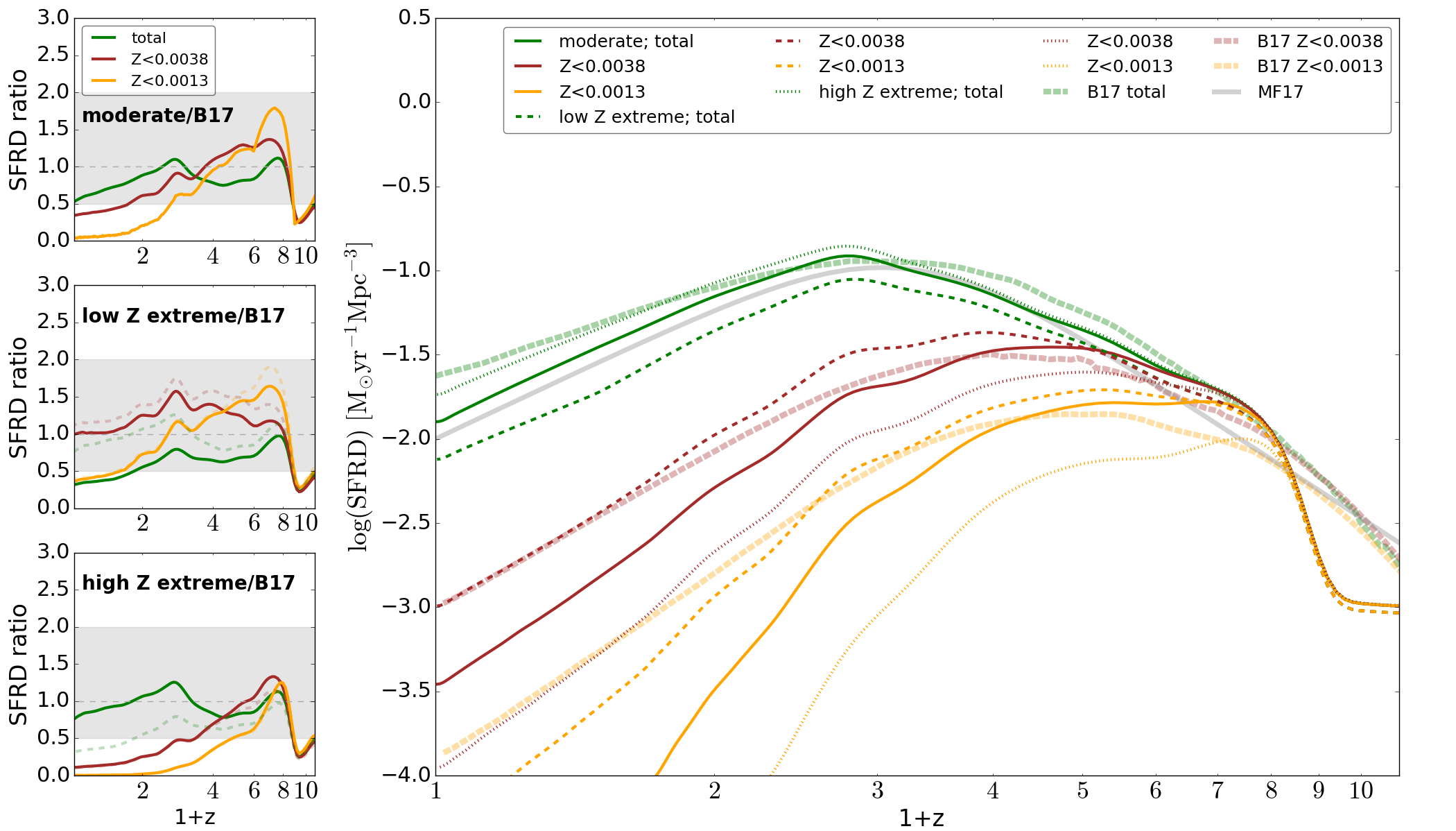}
\caption{ 
Comparison between the results from this study
and the results from the Illustris-1 cosmological simulations shown in \citet{Bignone17} (B17; see Fig. 1 therein).
The right panel shows the cosmic star formation rate density; total (green lines) 
and below certain metallicity thresholds:
Z$<$0.0038 (brown lines), Z$<$0.0013 (orange lines).
Thick dashed lines correspond to the data from B17, 
the solid, thin dashed and dotted lines show the results obtained for the
moderate, low Z extreme and high Z extreme variations respectively.
The thick gray solid line shows the cosmic SFRD from \citet{MadauFragos17} (M17) for the reference.
\newline
The left panels show the ratio of the SFRD(z) from our model to that from B17 for different
metallicity thresholds.
Different panels correspond to different model variations.
The gray band indicates a factor of 2 variation.
The faint dashed lines
show the effect of changing only the SFMR
(SFMR with no flattening in the middle and sharp flattening in the bottom panel).
The data from B17 has been converted to \citet{Kroupa01} IMF. 
Only the galaxies with stellar masses $M_{\ast}>10^{8} \Msun$ 
were considered in the comparison,
since this is the minimum  $M_{\ast}$ used by B17.
}
\label{fig: illustris_vs_model}
\end{figure*}

While the comparison with the predictions of cosmological simulations
cannot settle the correctness of our results, it is certainly interesting
to confront the two methods, especially in the regimes not covered by observations.
Here we briefly compare our results with the predictions of the large-scale Illustris-1
cosmological hydrodynamical simulations \citep{Vogelsberger14} presented in Fig 1. in \citet{Bignone17},
leaving a more detailed comparison with the simulations for future studies.
The comparison is shown in Fig. \ref{fig: illustris_vs_model}.
The right panel reproduces Fig. 1 from \citet{Bignone17} with additional
curves corresponding to the results from our model.
We show only three (out of six) lines presented in Fig. 1 in \citet{Bignone17}
to improve the readability of the plot.
Those are the lines corresponding to the total SFRD (\textit{no cut}
in \citealt{Bignone17}; green lines), the SFRD occurring below $Z$=0.0038 
(0.3$\Zsun$ cut in \citealt{Bignone17} $-$ $\Zsun=0.0127$ in their paper; brown lines)
and the SFRD occurring below $Z$=0.0013 (0.1$\Zsun$ cut in \citealt{Bignone17}; orange lines).
None of the curves corresponds to super-solar metallicity
(as \citealt{Bignone17} focus on long GRB host galaxies) 
and hence we cannot compare the distributions in that regime.
The comparison reveals that the closest match between the Illustris-1 results
and our model can be found for the low metallicity extreme variation
(the ratio stays within a factor of 2.5 up to $z\sim$8).
The ratio changes only slightly if a different SFMR flattening is used in combination
with the PP04 MZR 
(e.g. no flattening in SFMR instead of the sharp flattening used in the low-Z extreme
variation; compare solid and dashed lines in the middle panel on the left 
in Fig. \ref{fig: illustris_vs_model}).
The best agreement is reached for the $Z$=0.0038 metallicity cut.
The SFRD occurring below $Z$=0.0013 in this variation is underpredicted
with respect to the Illustris-1 simulations at  $z\lesssim$2 and overpredicted
at  $z\gtrsim$3, which suggests that the chemical evolution of the simulated
universe is slower than in our case.
In any case, at $z\sim$8 the SFRD from our model shows a steep decrease $-$ the
consequence of the evolution of the GSMF $-$ not reflected in the results from simulations.
The Illustris-1 simulations predict higher star formation rate density
at $z\lesssim$2 than our model and also more low metallicity star formation
up to $z\sim$5 and above z$\sim$8 for the moderate and high-Z extreme variations.
\\
We note that the results of the comparison would likely change for the
updated IllustrisTNG simulations \citep[e.g.][]{Pillepich18,Torrey19}, 
as one of the primary differences with respect to Illustris-1
is connected to the mass-metallicity relation from the simulations \citep{Torrey19}.
\citet{Torrey19} report the improved agreement with the observed MZR,
e.g. the flattening of the relation at high masses, which was not
reproduced in the previous version \citep{Torrey14} is now accounted for.
Also, the slope of the simulated MZR has changed.
The normalization of the MZR found in IllustrisTNG continuously decreases
with redshift, with the average rate at $z>2$ of $\approx -0.064$ dex per $\Delta z$=1.
  We note that this rate is much lower than assumed in our model based on the 
   observed evolution between $z$=2.2 and $z$=3.5
  (see Sec. \ref{sec: MZR: evolution} and \ref{sec: method: MZR}).

\subsection{Our results in the context of previous studies} \label{sec: discussion: comparison}

\begin{figure}
\centering
\includegraphics[scale=0.39]{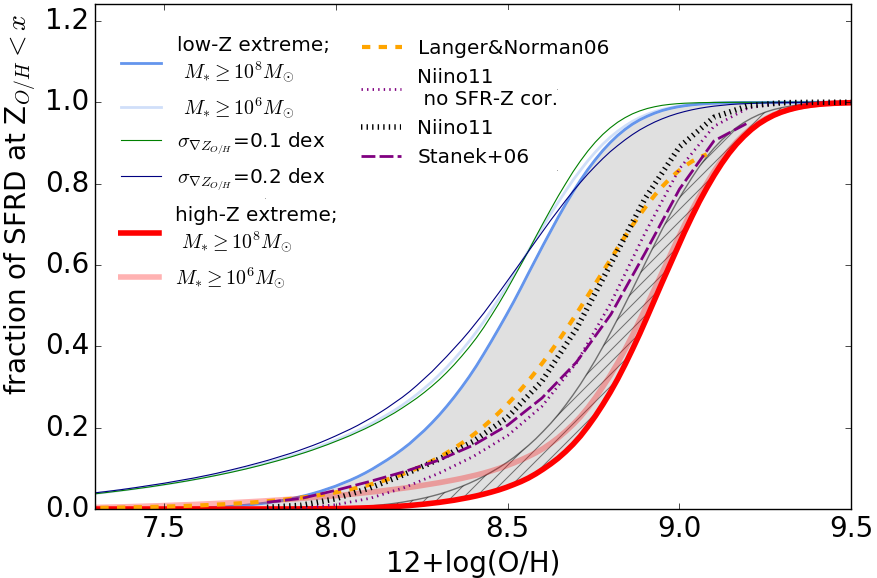}
\caption{
Cumulative fractions of total star formation at z$\sim$0 happening below a certain metallicity.
The solid lines show the results from our study 
(thick $-$ high-Z extreme, thin $-$ low-Z extreme) using galaxies with 
$M_{*}\geq 10^{8} \Msun$ (dark lines) or $M_{*}\geq 10^{6} \Msun$ (faint lines).
The gray area spans between the two extremes for the $M_{*}\geq 10^{8} \Msun$
mass cut.
The thin lines around the left-most extreme 
demonstrate the effect of varying the metallicity dispersion ($\sigma_{\nabla O/H}$) 
within galaxies between 0.1 dex (steeper curve) and 0.2 dex (flatter curve).
The hatched region on the right demonstrates the effect of changes in the SFMR only
(the right edge $-$ no flattening, the left edge $-$ sharp flattening in the SFMR).
\newline
We overplot the results from \citet{Stanek06}, \citet{Niino11} and based on the method from \citet{LangerNorman06}.
Despite the different assumptions, those results fall between the range indicated by the extreme cases
identified in our study.
}
\label{fig: CDF_compare}
\end{figure}

Several other studies used empirical scaling relations for star forming galaxies
to obtain the information about the distribution of the star formation
(or the formation of transients related to massive stars) across different metallicities.
Below we highlight the similarities and differences with respect to our method.
The details of the different methods are summarized in Tab. \ref{tab: others} in appendix \ref{app: others}.
We note that none of those studies set finding the SFRD(Z,z) distribution
as its main goal and only a very limited comparison of our results with
the published material is possible. 
\\ \newline
\citet{Stanek06} and \citet{Niino11} limited their analysis to low redshifts
($z\lesssim$0.3) and focused on application to long GRBs. 
Both studies take into account the intrinsic scatter in the SFMR and MZR.
\citet{Stanek06} does not take into account the correlation between the SFR and metallicity
due to FMR, while \citet{Niino11} shows a variation
in which they use the FMR from \citet{Mannucci11} instead of combining the SFMR and MZR.
\citet{Stanek06} consider two variations of the high-mass end slope of the SFMR -
relation with a downturn (also used by \citealt{Niino11}, see Fig. 1 therein)
and relation with flattening (similar to our sharp flattening case).
We note that such a downturn has not been observed in any of the subsequent studies
of the SFMR. This assumption may underestimate the star formation
happening at high metallicities.
\\ \newline
\citet{Dominik13} and \citet{LangerNorman06} use a slightly different approach than presented in this study:
they first obtain the metallicity distribution at each redshift (convolving MZR and GSMF)
and scale it with SFRD(z) to obtain the SFRD(Z,z)
 \footnote{ 
See also the study by \citet{Neijssel19} released during the revision process of this manuscript. 
We do not discuss their method here, but include it in the summary shown in Tab. \ref{tab: others}.
}.
Those authors constructed the $SFRD(Z,z)$ that can be extended to 
high redshifts, albeit with various simplifications.
One of them is that the adopted mass metallicity relation assumes no scatter so that there is a unique
correspondence between mass and metallicity, and the redshift evolution only affects the normalization of the relation. 
Furthermore, both methods implicitly assume a constant specific star formation rate 
(i.e. $SFR(M_{\ast})\propto M_{\ast}$; this is similar to our 'no flattening' case but with a different slope)
at all redshifts. 
They do not account for the FMR. \\
\citet{Dominik13} use MZR with the functional from \citet{Tremonti04} and redshift-dependent normalization,
adjusted to produce average metallicity of galaxies at $z=0$ of $Z$=0.03
or 0.016 in their \textit{high-end} and \textit{low-end} model respectively
(see Fig. 2 therein). 
\citet{LangerNorman06} use a non-evolving GSMF and 
assume that the MZR is a single power law (without the observed flattening at high $M_{*}$)
and its normalization decreases with redshift (at $z\gtrsim$2
the decrease is much slower than in any variation considered in our study).\\
Note that the assumed lack of flattening in MZR (and possibly SFMR) does likely not affect the results of studies
that use only the low-metallicity part of the distribution (as in some application to long GRB e.g. \citealt{LangerNorman06}),
but may be important for other applications 
(e.g. the model by \citealt{LangerNorman06} has been recently used by to calculate 
the merger rates of double compact objects in \citealt{Barrett18,Eldridge19}, 
this was also the application of the model by \citealt{Dominik13}).
\\ \newline
Another example of using scaling relations to infer the information about metallicity 
comes from \citet{MadauFragos17}. 
However, the authors give only the mass-averaged mean gas-phase metallicity
and not the entire distribution.
They use the mass-metallicity relation from \citet{Zahid14} and GSMF from several studies
to estimate the mean at different redshifts, assuming that the relation given by \citet{Zahid14}
can be extrapolated to z$\sim$7.
The resulting mass-averaged mean metallicity is overplotted in Fig. \ref{fig: SFRD_ZZ_z_103f_3} 
showing our moderate variation. 
This mean is very close to the peak metallicity in the high metallicity extreme
(note that the MZR assumed in \citet{MadauFragos17} is based on KK04 calibration,
as in the high-Z extreme case)
from our study at $z\lesssim$3 and is significantly higher at higher redshifts.
\\ \newline
Using the results from \citet{LangerNorman06},
we construct the cumulative distribution of total star formation
below a certain metallicity for their method and plot it in Fig. \ref{fig: CDF_compare}
together with the ones from \citet{Stanek06} (Fig. 2 therein, thick solid line\footnote{
The cumulative distribution for the case with SFMR that flattens at high masses 
discussed in \citet{Stanek06} is not shown and hence we only compare with their
default case involving SFMR with a downturn at high masses.
}), \citet{Niino11} (Fig. 4 and 5 therein, solid lines) and resulting from our model.
It can be seen that all the curves fall within the broad range defined by the extreme cases
from our study (gray area).
The hatched region shows how much the result would change only due to changes in the SFMR
on the example of the high-Z extreme model variation.
The width of this range is much smaller than the width of the gray area,
which shows that it is mostly set by the differences in the MZR.
The results from \citet{Niino11} were obtained for the same metallicity calibration 
as assumed in the high-Z extreme case.
They reveal a noticeably greater offset towards low metallicities
from the high-Z extreme curve than indicated by the hatched region, 
especially if the correlation between the SFR and $Z_{O/H}$
is accounted for (as in our model).
This difference may be (partially) explained by the high-mass turnover
in the SFMR used by \citet{Niino11}, although as discussed earlier in this section,
there are also other differences in the assumptions
between our study and \citet{Niino11}.
\\
Based on the data shown in the bottom panel of Fig. 2 from \citet{Dominik13}, we conclude that
their low-end model would fall well within the gray area shown in Fig. \ref{fig: CDF_compare}.
However, the high-end model from \citet{Dominik13} predicts hardly any $z\sim$0
star formation below $Z_{O/H}\sim$8.7 (see the middle panel in Fig. 2 therein) and therefore the 
cumulative distribution of the total star formation below certain $Z_{O/H}$
for that model would be shifted to noticeably higher metallicities with respect to our high-Z 
extreme shown in Fig. \ref{fig: CDF_compare}.

\section{Conclusions}
We combine empirical scaling relations for star forming galaxies
over wide range of masses and redshifts and other observational properties
of those galaxies to address the question
of (i) how was the cosmic star formation rate density distributed over metallicities
throughout the history of the Universe (SFRD(Z,z)) and
(ii) how uncertain is that distribution
given the currently unresolved problems
concerning the observationally inferred scaling relations.
\\
To evaluate the uncertainty of the SFRD(Z,z) obtained with our method,
we use different versions of the scaling relations, covering the range of
possibilities present in the literature.
We use four different versions of the mass $-$ metallicity relations
(based on different metallicity calibrations; 
with different normalizations, shapes and evolution with redshift),
vary the assumptions about the high mass end of the star formation $-$ mass relation
(assuming either a single power law, broken power law or sharp flattening at the high mass end)
and about the low mass end of the galaxy stellar mass function
(assuming a fixed low mass end slope or a slope that steepens with redshift).
We take into account the scatter present in the relations,
the observed correlation between the SFR and metallicity 
(the fundamental metallicity relation)
and the distribution of gas metallicity within galaxies.
We extend the analysis up to $z\sim$10, but we note that the MZR is not constrained by
observations above $z\sim$3.5 and hence our results at higher redshifts rely on extrapolations.
To quantify the differences between the variations of our model, we compare the 
fractions of stellar mass formed at low (below 10\% solar) and high (above solar) metallicity
at $z\leq$0.5, 3 and 10. 
Our main results can be summarized as follows:
\begin{itemize}
 \item 
We identify two variations $-$ the low and high metallicity extremes $-$ that
maximize the fraction of stellar mass formed at low and high metallicity respectively.
Those extremes are reached by combining the $PP04$ MZR 
(based on the \citealt{PettiniPagel04} O3N2 metallicity calibration)
with the SFMR with sharp flattening at high masses in the low-Z extreme case
and $KK04$ MZR 
(based on the \citealt{KobulnickyKewley04} metallicity calibration)
with the SFMR with no flattening at high masses in the high-Z extreme.
The fraction of mass formed in stars at low (high) metallicity since $z$=3 
differs by $\sim$18\% ($\sim$26\%) between the two extremes.
Those two variations also lead to different total stellar mass formed at each redshift
(with the higher stellar mass formed in the high-Z extreme).
This reduces the difference between the extremes if the stellar mass (instead of the fraction) 
formed at low metallicity is considered and
increases the difference between the amount of stellar mass formed at high metallicity.
\item
We find that the variations with the low mass end slope of the GSMF that steepens with redshift
significantly overpredict the total star formation rate density at $z\gtrsim$4
with respect to other observational estimates. The cause of this discrepancy is not clear.
Those variations were not considered at $z\gtrsim$4.
\item
The differences between the MZRs obtained for different metallicity calibrations
(in particular the differences in normalization) 
are the main cause of uncertainty in our results at $z\lesssim$4.
\item 
We compare the results from our model with the local specific core-collapse supernovae (CCSN) rate
as a function of metallicity from \citet{Graur17}.
The uncertainties in the CCSN-location metallicity estimates, as well as in the
translation of the total SFR to the number of CCSN do not allow us to draw any 
strong conclusions from that comparison.
However, it might suggest that the variations with the sharp flattening 
at the high mass end of the SFMR underestimate the local high metallicity star formation.
The other variations of our model show a reasonably good agreement with the data.
\end{itemize}
Our model is publicly available and can be used in studies
focusing on the properties of populations of systems composed of stars and their remnants,
stellar-evolution related transients and their likely environments.
In particular, it can be applied to calculate the rates of those transients and to evaluate
their uncertainty due to the assumed distribution of the cosmic star formation rate density 
across metallicities and redshifts.
It can also serve for comparison for the cosmological simulations.

\section*{Acknowledgements}
We would like to thank Leindert Boogaard, S{\o}ren Larsen, Svea Hernandez, Tereza Je{\v{r}}{\'a}bkov{\'a}, Jorryt Matthee, David Carton
and especially Jarle Brinchmann for valuable discussions. 
We also thank Or Graur, Zhiqiang Yan and the anonymous referee for their comments and Or Graur and Lucas Bignone
for sharing the data used in Fig. 11 and Fig. 12 respectively.
MC and GN acknowledge support from the Netherlands Organisation for Scientific Research (NWO).


\bibliographystyle{mnras}
\bibliography{SFRD_Z_z_bibliography} 



\appendix

\section{Metallicity $-$ nomenclature and solar composition}\label{app: nomenclature}
The term `metallicity` has been used variously in the literature,
depending on the object of interest and the method used.
It is generally understood as a measure of
the abundance of metals (elements heavier than helium) in a certain object or system
and defined as a fraction of total baryonic mass contained in metals:
\begin{equation}
 Z\equiv \frac{M_{metals}}{M_{baryons}}
\end{equation}
although this definition is mostly applicable in theoretical studies.
Observationally metallicity is often expressed in terms of the of the oxygen to hydrogen 
abundance ratio:
\begin{equation}
 Z_{O/H} \equiv 12 + \rm log \left(O/H \right) \equiv 12 + log\left( n_{O} / n_{H} \right)
\end{equation}
where $n_{i}$ stands for the number density of either oxygen or hydrogen.
This definition is commonly used e.g. in interstellar medium studies, since
oxygen is the most abundant heavy element and it is relatively easy to observe
due to its strong optical lines. The common assumption is that the other elements
scale linearly with the measured $Z_{O/H}$ maintaining the solar abundance ratios
(so that the conversion between the two measures has the form:
$log\left(Z/Z_{\odot}\right) = Z_{O/H} - Z_{O/H \odot}$, where $Z_{\odot}$ and $Z_{O/H \odot}$
stand for the solar metallicity in both measures).
\\
However, the exact value of solar metallicity and solar composition is presently unknown. 
For instance, a widely used result obtained by \citet{Asplund09}
based on 3D modeling of the solar atmosphere gives $Z_{\odot}$=0.0134 
and $Z_{O/H \odot}$=8.69.
 Such low values of metallicity made it difficult
 to recover the helioseismically inferred information about the solar interior,
 such as sound speed profile, the location of the convective zone boundary and surface helium abundance
 \citep[e.g.][]{Serenelli09,Serenelli16}.
 In fact, the metallicity estimates based on helioseismology are systematically higher,
  with $Z_{O/H \odot}\sim$8.85 \citep[e.g.][]{DelahayePinsonneault06,Villante14}.
 Using the abundances found by \citet{Villante14}, the resulting $Z_{\odot}\approx$0.019.
 Similarly high values were obtained by the earlier solar atmosphere studies, 
 e.g. \citet{AndersGrevesse89} found $Z_{O/H \odot}$=8.83
and \citet{GrevesseSauval98} found $Z_{O/H \odot}$=8.93.
Asplund et al. (2009) recalculated $Z_{\odot}$ from those studied using the updated He abundances
 from helioseismology and found $Z_{\odot}$=0.017 and $Z_{\odot}$=0.0201 respectively.

\section{Initial mass function - corrections}\label{app: method: IMF}
The choice of a particular IMF impacts the estimates of both stellar masses and star formation rates
inferred from observations and as a result affects all of the galaxy scaling relations used in our model.
The results obtained assuming a certain IMF can be
translated to another IMF by shifting the mass and SFR estimates by a 
certain value (the change is in the same direction, but generally with different magnitude).
To correct for differences between the IMFs used in various studies 
(either \citet{Salpeter55}, \citet{Kroupa01} or \citet{Chabrier03}) 
we use the correction factors listed in table \ref{tab: IMF_corrections}, unless explicitly stated otherwise.
These include the mass conversions as given in \citet{Speagle14} (see eq. 2 therein)
and simple SFR correction calculated as the ratio of the total mass in both IMFs 
assuming the same number of massive stars 
\citep[$>$10 $\Msun$ $-$ stars that are responsible for the bulk of
the measured UV light; e.g.][]{Boogaard18,Fermi18}. 

 \begin{table}
\centering
\small
\caption{
  conversion factors used in this study to correct for different IMFs used in various studies;
  \citet{Kroupa01}: K01, \citet{Chabrier03}: Ch03, \citet{Salpeter55}: Sal. \newline
  Mass correction from IMF1 to IMF2: $\rm log\left( M_{IMF2} \right)=log\left( M_{IMF1} \right)+\Delta log(M)$
  \newline
  SFR correction from IMF1 to IMF2: $\rm SFR_{IMF2}=SFR_{IMF1}\times K_{IMF}$
}
\begin{tabular}{c c c c }
\hline
 conversion factor & K01 to Sal & K01 to Ch03 &  \\ \hline
  \hline
  $\Delta$log(M) & +0.21 dex & $-$ 0.03 dex &\\ \hline
  $K_{IMF}$ & 1.5 & 0.94 & \\ \hline
\end{tabular}
\label{tab: IMF_corrections}
\end{table}

\section{Metallicity of stars - differences between the variations}
\subsection{The metallicity at which the most of the stellar mass was formed}\label{app: peak}

\begin{figure*}
\centering
\includegraphics[scale=0.46]{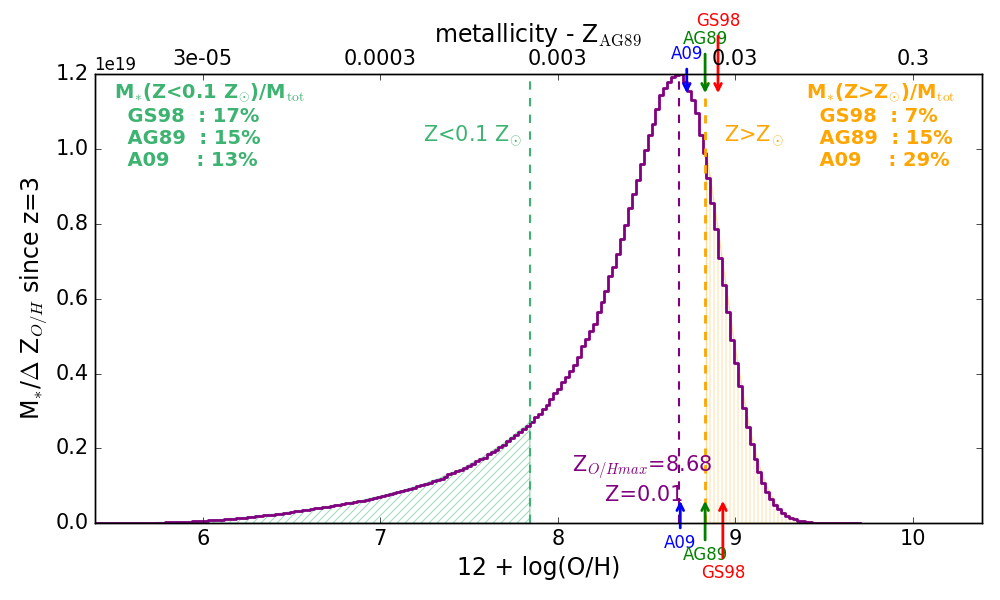}
\caption{ 
Distribution of the mass formed in stars since redshift z=3 at different metallicities 
(measured in oxygen to hydrogen abundance 12 + log(O/H) $-$ bottom scale, or as a metal mass fraction $-$ top scale)
for the model assuming moderate flattening at the high mass end of SFMR, M09 mass metallicity relation and $\alpha_{fix}$=$-$1.45
(SFRD(Z,z) for that variation is shown in Fig. \ref{fig: SFRD_ZZ_z_103f_3}).
The green tail of the distribution corresponds to the stars forming at $Z<0.1 \ \Zsun$, while the orange one at $Z>\Zsun$.
The purple line indicates the peak of the distribution.
This variation of the model falls in between the extreme cases in terms of the fraction of low and high metallicity star formation.
The fractions of mass formed in stars at $Z<0.1 \ \Zsun$ and $Z>\Zsun$
for the discussed model are given in the top left and top right corner of the figure respectively, 
for the three commonly used solar metallicity estimates.
In this paper we assume \citet{AndersGrevesse89} solar metallicity (AG89; indicated with the green arrows) 
and the top scale was calculated from 12 + log(O/H)
using the solar abundances found by these authors. 
The other two commonly adopted solar metallicity estimates are 
indicated with blue \citep[A09;][]{Asplund09} and red \citep[GS98;][]{GrevesseSauval98}
arrows for the reference.
}
\label{fig: distr_103f_3}
\end{figure*}

\begin{figure*}
\centering
\includegraphics[scale=0.4]{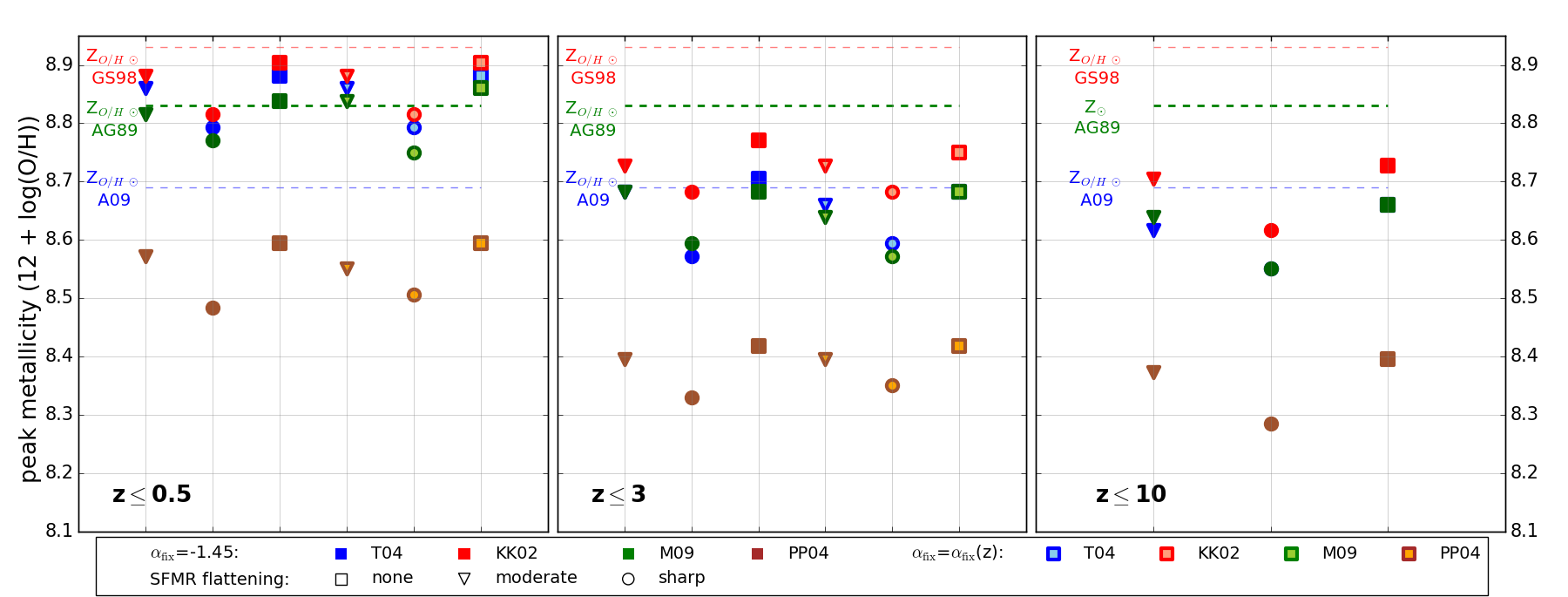}
\caption{ 
12 + log(O/H) at the peak of the distribution of mass formed in stars at different metallicities (see Figure \ref{fig: distr_103f_3})
since redshift z=0.5 (left), z=3 (middle) and z=10 (right) for different variations of the model.
Different symbols correspond to the assumptions about the high mass part of the SFMR: squares $-$ no flattening,
triangles $-$ moderate flattening, circles $-$ sharp flattening.
Colours distinguish different assumptions about the MZR and low mass end slope of the GSMF.
The dark colours correspond to $\alpha_{fix}$=$-$1.45, where the MZRs are: blue $-$ T04, red $-$ KK04, green $-$ M09, brown $-$ PP04;
The symbols with contours and light colours correspond to $\alpha_{fix}=\alpha_{fix}(z)$ steepening with redshift, where the MZRs are
light blue $-$ T04, light red $-$ KK04, light green $-$ M09, orange $-$ PP04.
We exclude the variations with $\alpha_{fix}(z)$ at high redshifts z$\gtrsim$4 (see text).
The solar oxygen abundance estimates from \citet{AndersGrevesse89} (AG89; green),
\citet{Asplund09} (A09; blue) and \citep{GrevesseSauval98} (GS98; red) are indicated with the horizontal dashed lines for the reference.
}
\label{fig: FOH_max}
\end{figure*}

As discussed in Sec. \ref{sec: results: differences}, 
in almost all cases discussed in our study most of the stellar mass forms outside the assumed
low and high metallicity regimes (i.e. between 0.1 $\Zsun$ and $\Zsun$).
Here we look at the exact location of the peak of the distribution of mass formed in stars across metallicities
for different redshift ranges and discuss its dependence on the assumptions about the MZR, SFMR and GSMF.
\\
Figure \ref{fig: distr_103f_3} shows an example of such distribution.
We note that for all the variations considered in our study
such distributions can be described by a combination of a left-skewed normal 
distribution and a power-law tail at low $Z_{O/H}$.
When higher redshifts are included, the peak of the skewed-normal part shifts towards lower metallicities
 and the distribution broadens on the low metallicity side while the tail builds up.
\\ \newline
\\
The peak metallicity for different model variations is shown in Fig. \ref{fig: FOH_max}.
It shows relatively little dependence on the assumed shape of the high mass part of the SFMR.
This can be seen by comparing at the location of the symbols with the same colour and different
shapes in Fig. \ref{fig: FOH_max}.
The difference is $\sim$0.1 dex. 
Comparing points plotted with the same symbol and colour but
with and without edge in Fig. \ref{fig: FOH_max} we conclude that
the effect of the low mass end slope of the GSMF is negligible.
To investigate the effect of the assumed mass metallicity relation 
(MZRs obtained for different metallicity calibrations),
we compare the locations of points plotted with the same symbol but with different colours.
It can be seen that for the T04, KK04 and M09 MZRs the difference is within 0.1 dex.
Those three relations have similar (high) normalizations at z$\sim$0.
Additionally, both the M09 and T04 relations are relatively steep and evolve with redshift at a similar rate
and hence the green and blue points corresponding to those relations can be found close to each other or 
they even overlap in Fig. \ref{fig: FOH_max}. The KK04 MZR is flatter than those two relations
(so that according to this MZR low mass galaxies have higher metallicities compared to M09 or T04)
and shows a milder redshift evolution and its peak metallicity is the highest.
The peak metallicity obtained for the PP04 MZR (with the low initial normalization) falls $\gtrsim$0.25 dex 
below the other three MZRs.
\\

\subsection{Low metallicity mass fraction} \label{app: lowZ mass fraction}
\begin{figure*}
\centering
\includegraphics[scale=0.4]{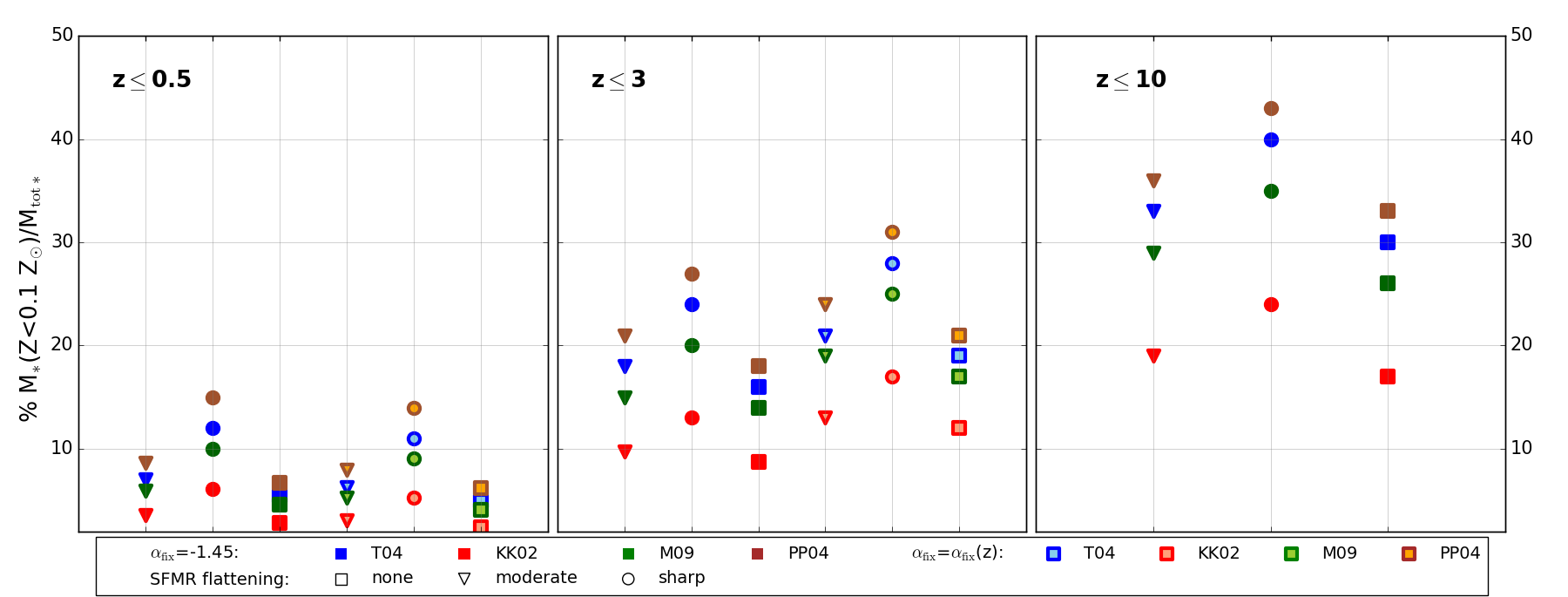}
\caption{ 
The fraction of mass formed in stars at low metallicities
\citep[$<0.1\Zsun$; calculated assuming solar metallicity from][]{AndersGrevesse89}
since redshift z=0.5 (left), z=3 (middle) and z=10 (right) for different variations of the model.
Different symbols correspond to the assumptions about the high mass part of the SFMR: squares $-$ no flattening,
triangles $-$ moderate flattening, circles $-$ sharp flattening.
Colours distinguish different assumptions about the MZR and low mass end slope of the GSMF.
The dark colours correspond to $\alpha_{fix}$=$-$1.45, where the MZRs are: blue $-$ T04, red $-$ KK04, green $-$ M09, brown $-$ PP04;
The symbols with contours and light colours correspond to $\alpha_{fix}=\alpha_{fix}(z)$ steepening with redshift, where the MZRs are
light blue $-$ T04, light red $-$ KK04, light green $-$ M09, orange $-$ PP04.
We exclude the variations with $\alpha_{fix}(z)$ at high redshifts z$\gtrsim$4 (see text).
}
\label{fig: lowZ}
\end{figure*}

The fraction of mass formed in stars at low metallicities since $z$=0.5, 3 and 10 for different variations 
of the model is shown in Fig. \ref{fig: lowZ}.
The choice of the MZR (or metallicity calibration) has a decisive role in setting
the fraction of the low metallicity star formation 
(compare the same symbols plotted with different colours).
As expected, this fraction is the highest for PP04 (brown) and the lowest for
the KK04 MZR (red) at all redshifts.\\
The second important factor is the flattening at the high mass end of the SFMR.
For a given MZR (fixed colour in Fig. \ref{fig: lowZ}) the sharp flattening case (circles)
leads to the highest fraction of mass formed at low metallicity
 and this fraction is the lowest if the SFMR is a single power-law (`no flattening` case; squares).
\\
The evolving low mass end slope of GSMF leads to a slight increase in the amount of mass
formed at low $Z_{O/H}$ since $z$=3 with respect to the fixed slope case 
(as there are more low mass galaxies forming low metallicity stars).
This variation is not considered at higher redshift for the reasons discussed in sec. \ref{sec: results: alpha_z}.
\\ \newline

\section{Comparison with core-collapse supernovae rates}
 
 \subsection{Cosmic SFRD vs CCSN rate density}\label{app: CCSN rates}
It is commonly assumed that the CCSN rate is proportional to the SFR and the
constant of proportionality depends solely on the IMF and can be
estimated as a number of stars that explode as CCSNe per unit mass 
\citep[the efficiency of formation of CCSN progenitors; e.g. equation 16 in][]{MadauDickinson14}.
This calculation requires assuming a certain initial mass range for the progenitors of CCSNe.
However, the exact limits of this mass range are not known 
\citep[and the range itself may in fact consist of so-called islands of explodability,
e.g.][]{SukhboldWoosley14,PejchaThompson15,Sukhbold16}.
Furthermore, those limits depend on the stellar metallicity \citep[e.g.][]{Heger03,Langer12, Doherty15}
and hence likely change throughout the cosmic history.
The expected CCSN progenitor mass range is further modified by binary interactions and stellar mergers
\citep[e.g.][]{Eldridge08,Zaparatas17}.
As argued by \citet{Sana12}, over 70\% of the most massive stars are born in interacting binary systems
and $\sim$20 $-$ 30\% will experience a merger with their companion.
Stellar interactions and mergers can act both to reduce the number of CCSN expected from a certain stellar population
(e.g. by mergers of two CCSNe progenitors) and to increase that number 
(e.g. by mergers of the more common low mass stars
that result in a CCSN progenitor). 
\citet{Zaparatas17} find that the overall effect of accounting for the binaries
is to increase the number of the CCSN expected to form from a stellar population of the same mass
(they find $\sim$14\% increase, although the exact number is model dependent).
Another complication added by binaries is the fact that 
for the same amount of star formation, stellar population containing binaries would lead to a
higher UV luminosity than the one consisting only of single stars and therefore would lead to an overestimate
in the SFR measured using the common UV-light based tracers 
\citep[e.g.][the magnitude of this effect varies depending on the SFR tracer and metallicity of the stellar population]{XiaoEldridge15}.
In other words, the actual number of stars born as CCSN progenitors would be smaller than what results from 
the inferred SFR, which would lead to an overestimate of the number of expected CCSN.
The combined effect of the presence of binaries in the observed stellar population on the 
CCSN formation efficiency is not clear.
The comparison is further complicated by the fact that the
observational CCSN rate estimates are not always consistent with each other, 
which may be partially attributed to different methods used to correct for the number of CCSN
missed due to obscuration by dust \citep[e.g.][]{Graur15}.
 
\subsection{Central metallicity vs metallicity at the supernova location}\label{app: SN location Z}

\citet{Galbany16} suggest that the true metallicity at the CCSN location should be contained within
the range reaching metallicity 0.1 dex lower than the central value.
However, the \citet{Marino13} metallicity calibration used in \citet{Galbany16} probes a relatively small metallicity range 
and as shown in \citet{Sanchez-Menguiano16}, leads to relatively flat metallicity gradient estimates.
For instance, \citet{Sanchez-Menguiano16} find the typical metallicity gradient of around $-$0.07 dex/r$_{e}$ 
using the \citet{Marino13} calibration, $-$0.11 dex/r$_{e}$ with the PP04 calibration and 
$-$0.14 dex/r$_{e}$ using the theoretical calibration from \citet{Dopita13}.
Hence, based on the metallicity gradient studies and taking this into account
the fact that CCSN are found to explode within 2 r$_{e}$ from the centre,
\citep[with an average distance of around 1 r$_{e}$, e.g.][]{Galbany16}, 
one can expect to find also larger offsets between the central and local metallicity than 0.1 dex proposed by 
\citet{Galbany16} (especially when using one of the theoretical metallicity calibrations).
\citet{Sanders12} find a median offset between the nuclear and explosion site metallicity for their stripped envelope
SN sample of $\sim$0.08 dex in the PP04 scale, although with a substantial scatter (see Fig. 12 therein). 
However, the galaxies considered in this study contain objects at higher redshift and with smaller angular sizes
than those used in \citet{Graur17}. Hence, the nuclear metallicity in those galaxies was measured
with a larger covering fraction (averaging information from a larger portion of the galaxy) 
and can be expected to be closer to the metallicity at the SN explosion site.
\citet{Modjaz11} compared the local metallicities of part of their sample of stripped envelope supernovae with the
central metallicities obtained from the nuclear SDSS spectra using the PP04 metallicity scale and found the average 
difference of 0.13 dex and in the extreme cases reaching up to 0.24 dex.
 
\subsection{Specific CCSN rate vs metallicity - dependence on the scatter in metallicity}\label{app: specific CCSN rate slope}

The shape of the dependence of the specific CCSN rate on metallicity is a consequence
of the scaling relations in star forming galaxies.
The specific star formation rate (SFR/$M_{\ast}$) decreases with mass, so that there are fewer CCSN per stellar mass
in more massive host galaxies. The mass - metallicity relation tells that with the increasing stellar mass 
(and hence decreasing specific SFR or specific CCSN rate) metallicity increases up to a certain $Z_{O/H}$ value,
at which the MZR flattens, hence the slope of the specific CCSN rate $-$ metallicity relation steepens around the 
metallicity at which the MZR begins to flatten. 
The presence of the scatter around the MZR pushes a fraction of the star formation
into higher and lower metallicity than what results from the MZR, 
but the anticorrelation between the SFR and metallicity 
(resulting from the fundamental metallicity relation)
reduces that `smearing` effect of the scatter and acts to further steepen the high-metallicity slope.
The gray solid line in Fig. \ref{fig: specific rate metallicity} shows the specific CCSN rate $-$ metallicity
relation when only the scatter in the MZR ($\sigma_{0}$) is taken into account.
It can be seen that the high-metallicity slope of the gray curve is steeper than in the case of the blue solid curve,
which corresponds to the same model variation, but taking into account the fact that the
stars within galaxies can form with a range of metallicities ($\sigma_{\nabla O/H}$=0.14 dex).
Hence, the assumed width of this distribution has a noticeable effect on the high metallicity end slope
and on the location of the turnover in the specific CCSN rate $-$ metallicity dependence.\\
For instance, the power law slope of the high metallicity Z$_{O/H}\gtrsim 8.5$ part of this relation
is $\sim$ $-$1 for $\sigma_{\nabla O/H}$=0.14 dex, $\sim$ $-$1.4 for $\sigma_{\nabla O/H}$=0.1 dex
and for $\sim$ $-$0.65 for $\sigma_{\nabla O/H}$=0.2 dex.

\section{Closer look at the secondary assumptions} \label{app: secondary assumptions}
\subsection{The distribution of the star formation over metallicities within galaxies}\label{app: gradients}
As discussed in Sec. \ref{sec: MZR: gradients}, there is a range of metallicities at which
stars form within their host galaxies. 
\\
Our treatment of the distribution of metallicity of the star formation within galaxies
is very simplified. We assume that it can be described as a normal distribution centered
at the metallicity resulting from the MZR 
(taking into account the anticorrelation between the metallicity and the SFR) and with a
dispersion $\sigma_{\nabla O/H}$=0.14 dex for all galaxies and redshifts and 
irrespective of the choice of the MZR (metallicity calibration).\\
However, the range of metallicities found in the HII regions (setting the width of the distribution),
as well as the value of the characteristic metallicity gradient depend
on the assumed metallicity calibration \citep[e.g.][]{Sanchez-Menguiano16}.
The value of $\sigma_{\nabla O/H}$ that we use is based on the averaged results obtained using
the \citet{Marino13} and the \citet{PettiniPagel04} metallicity calibrations and could be higher if different 
(theoretical) metallicity calibrations were used.
\\
On the other hand, the offsets between
the metallicities measured at the location of the CCSN and the central metallicity of
its host galaxies (which also indicate the range of metallicities at which the stars form
within galaxies) reported in the studies using the empirical metallicity calibrations mentioned above
are smaller than 2$\times \sigma_{\nabla O/H}$ (see appendix \ref{app: SN location Z} and references therein).
\\
Furthermore, not all galaxies reveal a single slope negative metallicity gradient.
\citep{Carton18} studied galaxies at intermediate redshifts and found a significant fraction
of galaxies showing gradients that are consistent with being flat and
several cases of positive gradients, which might suggest that the simple common negative metallicity gradient
found in the galaxies in local Universe may not be relevant at higher redshifts.
\\
Finally, the detailed studies of metallicity gradients 
within galaxies (on which we based our assumptions) are limited to the relatively local Universe.
It is thus relevant to ask how much our results depend on this particular assumption.
\subsubsection{The importance for our results}
To asses the importance of the assumptions about the distribution of the star formation over metallicities
within galaxies on our results, we vary the value of $\sigma_{\nabla O/H}$ in the range 0.1$-$0.2 dex.
The effect on the results (summarized in Table \ref{tab: Mtot}) is minor:
the difference in the total stellar mass formed and the fraction of the low metallicity star formation
is negligible and the fraction of the high metallicity star formation differs by less than 3\%
between the calculations with $\sigma_{\nabla O/H}$=0.1 and 0.2 dex for all model variations.
The distribution of mass formed in stars at different metallicities since a certain redshift 
(see Fig. \ref{fig: distr_103f_3}) slightly broadens as $\sigma_{\nabla O/H}$ increases,
however this broadening only affects the low metallicity (Z$<0.1 \Zsun$) tail of the distribution 
at high redshifts $z>$5.
At the same time the peak metallicity shifts to lower values 
(by less than 0.03 dex for the distribution at z$<$0.5 and less than 0.09 dex at z$<$10 for all model variations), 
which reduces the impact of the broadening on the fraction of the high metallicity star formation.
However, as discussed in Sec. \ref{sec: CCSN rates}, changing $\sigma_{\nabla O/H}$ noticeably affects
the slope of the high metallicity end of the specific CCSN rate $-$ metallicity relation
(see Fig. \ref{fig: specific rate metallicity} and Sec. \ref{app: specific CCSN rate slope}).

\subsection{Contribution from passive galaxies}\label{app: passive}

In this study we use the scaling relations and galaxy mass functions obtained for star forming galaxies.
However, various criteria to select those galaxies are used in the literature and
a fraction of galaxies classified as passive may still have some residual star formation.
The fraction of passive galaxies increases towards lower redshifts \citep[e.g.][]{Muzzin13,Davidzon17}
and hence their contribution to the global star formation budget can be expected to be
the highest in the local Universe.
\citet{RenziniPeng15} use the SDSS DR7 release data to study the star formation $-$ mass relation of
the local galaxies without imposing any cuts on the star formation within galaxies in their sample, 
i.e. including quenched galaxies.
They find that a low level of star formation is still present within the passive population 
(although they note that in many cases the measured SFR may be an upper limit)
and show that while the majority of mass in the local Universe reside in those galaxies, 
they do not contribute significantly to the ongoing star formation.
\subsubsection{Importance for our results}
To obtain a rough estimate of the upper limit on the local star formation rate density
contributed by passive galaxies,
we construct the `upper limit star formation $-$ mass relation` for passive galaxies by 
fitting a line connecting the two low SFR peaks in the SFR $-$ stellar mass plane shown in Fig. 4 in \cite{RenziniPeng15}.
Combining it with the local galaxy stellar mass function for the passive (red)
population of galaxies from \citet{Baldry12} and integrating over the stellar masses between
$10^{8} - 10^{12} \Msun$, we find that the contribution of passive galaxies to the local SFRD
is more than a factor of 80 smaller than the SFRD obtained for the star forming galaxies in that mass range
in the moderate variation of our model.
Most of this residual star formation occurs in massive galaxies and hence adds to the high metallicity
tail of the SFRD(Z,z) distribution.
Assuming that all the star formation from passive galaxies adds to the star formation at Z$>\Zsun$
and their contribution to the global SFRD is at the level corresponding to the estimated upper limit,
the fraction of mass formed in stars at high metallicity at $z<0.5$ in the moderate variation of our model
would increase by $\lesssim$1\%.
Taking into account the fact that the fraction of passive galaxies decreases with redshift, 
only some of them show some residual star formation and the above estimate used the upper limit on the
level of that residual star formation rate, we conclude that the omitted star formation in passive
galaxies has negligible effect on our results.

\section{SFRD(Z,z) from scaling relations - other studies}\label{app: others}

Table \ref{tab: others} summarizes the assumptions made by several different authors using empirical
(or mixed empirical and theoretical/simulated) 
scaling relations to infer SFRD(Z,z).

 \begin{table*}
\centering
\small
\caption{
Summary of the assumptions made by different authors using empirical
(or mixed empirical and theoretical/simulated e.g. \citealt{Dominik13}, \citealt{Neijssel19}) 
scaling relations to infer SFRD(Z,z). In the row 'method' we indicate whether the authors 
combine SFMR, MZR \& GSMF (as in our study) or convolve MZR \& GSMF to obtain metallicity distribution (as a function of $z$)
and obtain the scaling using SFRD(z).
\newline
'+': included as a variation\newline
$^i$: \citet{Dominik13} use MZR with the functional from Tremonti et al. (2004) and redshift-dependent normalization. 
The normalization is calculated based on the results from Pei et al. (1999) and Young \& Fryer (2007) 
and subsequently adjusted to produce average metallicity of galaxies at z = 0 of 1.5 $\Zsun$ or 0.8 $\Zsun$ in
their high-end and low-end model respectively (see Fig. 2 therein, note that $\Zsun$=0.02 in Dominik et al. 2013).
\newline
$^{ii}$: see Fig. 1 in \citet{Niino11}
\newline
$^{iii}$: in their fiducial model; the distribution of metallicity at each redshift is described as a log-normal distribution
	  with $\sigma$=0.39 dex
\newline
$^{iv}$: \citet{Furlong15} - fit to $z$ evolution (cosmological simulations); \citet{Panter04} - non-evolving
}
\begin{tabular}{c c c c c c}
\hline
 study: & \citet{Stanek06} & \citet{Niino11} & \citet{LangerNorman06} & \citet{Dominik13} & \citet{Neijssel19} \\ \hline
 method: & SFMR, MZR \& GSMF & SFMR, MZR \& GSMF & SFRD(z), MZR \& GSMF & SFRD(z), MZR \& GSMF & SFRD(z), MZR \& GSMF \\ \hline
 $z$ & $\lesssim$0.25 & $\lesssim$ 0.3 & $\lesssim$10 & $\lesssim$10 & $\lesssim$6 \\ \hline
  log($M_{*}/\Msun$) & $>$7.4 & $>$8 & - &  $>$7 & - \\ \hline
 \multirow{3}{*}{MZR}& 	   &      &                  &                  & +\citet{Savaglio05} \\
		     &T04 & KK04 & single power law & modified T04$^i$ & +\citet{Ma16} \\ 
		     &     &      &                  &                  & +\citet{LangerNorman06}\\\hline
 \multirow{2}{*}{MZR(z)} & - & - & normalization & decreasing  & normalization \\
 & - & - & -0.15 dex per $\Delta z$=1 & normalization$^i$ & -0.23 dex per $\Delta z$=1$^{iii}$ \\ \hline
 $\sigma_{\rm MZR}$ & 0.1 dex & 0.1 dex & - & - & - / 0.39 dex$^{iii}$ \\ \hline
 \multirow{3}{*}{GSMF} & \citet{Bell03} & \citet{Bell03} & \citet{Panter04} & \citet{Fontana06} & +\citet{Panter04} \\ 
    &             &              &              & (fit to $z$ evolution)                 &+\citet{Furlong15} \\
   & non-evolving & non-evolving & non-evolving & non-evolving at $z>$4 & various$^{iv}$\\ \hline
 \multirow{3}{*}{SFMR} & \citet{Brinchmann04} & \citet{Brinchmann04} & $\sim$const. specific SFR & $\sim$const. specific SFR & $\sim$const. specific SFR\\ 
		      &  (downturn at high M$_{*}$)$^{ii}$ &  (downturn at high M$_{*}$)$^{ii}$ & SFR(M$_{*}$)$\propto$M$_{*}$ & SFR(M$_{*}$)$\propto$M$_{*}$ & SFR(M$_{*}$)$\propto$M$_{*}$\\
		      & +: sharp flattening & & & & \\ \hline
 $\sigma_{SFR}$	& 0.3 dex & 0.3 dex & - & - & - \\ \hline
 FMR & - & +: \citet{Mannucci11} & - & - & - \\ \hline
 $\sigma_{\nabla Z_{O/H}}$ & - & +0.1/0.3/0.5 dex & - & - & - \\ \hline
 
\end{tabular}
\label{tab: others}
\end{table*}

\bsp	
\label{lastpage}
\end{document}